\documentclass[sigconf,natbib=true]{acmart}
\AtBeginDocument{%
  \providecommand\BibTeX{{%
    \normalfont B\kern-0.5em{\scshape i\kern-0.25em b}\kern-0.8em\TeX}}}
\copyrightyear{2023}
\acmYear{2023}
\setcopyright{rightsretained}
\acmConference[CIKM '23]{Proceedings of the 32nd ACM International
Conference on Information and Knowledge Management}{October 21--25,
2023}{Birmingham, United Kingdom}
\acmBooktitle{Proceedings of the 32nd ACM International Conference on
Information and Knowledge Management (CIKM '23), October 21--25, 2023, Birmingham, United Kingdom}
\acmDOI{10.1145/3583780.3614923} 
\acmISBN{979-8-4007-0124-5/23/10}
\usepackage[ruled]{algorithm2e}
\usepackage{amsmath}
\usepackage{color}
\usepackage{subfig}
\usepackage{balance}
\usepackage{ifthen}
\usepackage{multirow}
\usepackage{enumitem}
\usepackage{pifont}
\usepackage{bbding}
\usepackage{wasysym}
\usepackage{ulem}
\usepackage[english]{babel} 
\usepackage{makecell}
\usepackage{tabularx}
\usepackage[T1]{fontenc}
\usepackage[utf8]{inputenc}
\usepackage{babel}
\usepackage[font=small,labelfont=bf,tableposition=top]{caption}
\usepackage{booktabs}
\usepackage{threeparttable}
\newcommand{\M}{\textrm{I}^3}
\definecolor{exampleblue}{RGB}{30,144,255}
\definecolor{examplepink}{RGB}{255,192,203}
\definecolor{exampleyellow}{RGB}{255,227,132}
\begin{document}
\title{$\M$ Retriever: Incorporating Implicit Interaction in Pre-trained Language Models for Passage Retrieval}
\author{Qian Dong}
\email{dq22@mails.tsinghua.edu.cn}
\affiliation{%
  \institution{DCST, Tsinghua University\&}
  \institution{Quan Cheng Laboratory\&}
  \institution{Zhongguancun Laboratory}
  \city{Beijing}
  \country{China}}
\author{Yiding Liu}
\email{liuyiding.tanh@gmail.com}
\affiliation{%
  \institution{Baidu Inc.}
  \city{Beijing}
  \country{China}}
\author{Qingyao Ai}\authornote{Corresponding author.}
\email{aiqy@tsinghua.edu.cn}
\affiliation{%
  \institution{DCST, Tsinghua University\&}
  \institution{Quan Cheng Laboratory\&}
  \institution{Zhongguancun Laboratory}
  \city{Beijing}
  \country{China}}
\author{Haitao Li}
\email{liht22@mails.tsinghua.edu.cn}
\affiliation{%
  \institution{DCST, Tsinghua University\&}
  \institution{Quan Cheng Laboratory\&}
  \institution{Zhongguancun Laboratory}
  \city{Beijing}
  \country{China}}
\author{Shuaiqiang Wang}
\email{shqiang.wang@gmail.com}
\affiliation{%
  \institution{Baidu Inc.}
  \city{Beijing}
  \country{China}}
\author{Yiqun Liu}
\email{yiqunliu@tsinghua.edu.cn}
\affiliation{%
  \institution{DCST, Tsinghua University\&}
  \institution{Quan Cheng Laboratory\&}
  \institution{Zhongguancun Laboratory}
  \city{Beijing}
  \country{China}}
\author{Dawei Yin}  
\email{yindawei@acm.org}
\affiliation{%
  \institution{Baidu Inc.}
  \city{Beijing}
  \country{China}}
\author{Shaoping Ma}
\email{msp@tsinghua.edu.cn}
\affiliation{%
  \institution{DCST, Tsinghua University\&}
  \institution{Quan Cheng Laboratory\&}
  \institution{Zhongguancun Laboratory}
  \city{Beijing}
  \country{China}}
\renewcommand{\shortauthors}{Qian Dong et al.}
\begin{abstract}
Passage retrieval is a fundamental task in many information systems, such as web search and question answering, where both efficiency and effectiveness are critical concerns. In recent years, neural retrievers based on pre-trained language models (PLM), such as dual-encoders, have achieved huge success. Yet, studies have found that the performance of dual-encoders are often limited due to the neglecting of the interaction information between queries and candidate passages. Therefore, various interaction paradigms have been proposed to improve the performance of vanilla dual-encoders. Particularly, recent state-of-the-art methods often introduce late-interaction during the model inference process. However, such late-interaction based methods usually bring extensive computation and storage cost on large corpus. Despite their effectiveness, the concern of efficiency and space footprint is still an important factor that limits the application of interaction-based neural retrieval models. To tackle this issue, we $\textbf{I}$ncorporate $\textbf{I}$mplicit $\textbf{I}$nteraction into dual-encoders, and propose $\textbf{I}^3$ retriever. In particular, our implicit interaction paradigm leverages generated pseudo-queries to simulate query-passage interaction, which jointly optimizes with query and passage encoders in an end-to-end manner. It can be fully pre-computed and cached, and its inference process only involves simple dot product operation of the query vector and passage vector, which makes it as efficient as the vanilla dual encoders. We conduct comprehensive experiments on MSMARCO and TREC2019 Deep Learning Datasets, demonstrating the $\M$ retriever's superiority in terms of both effectiveness and efficiency. Moreover, the proposed implicit interaction is compatible with special pre-training and knowledge distillation for passage retrieval, which brings a new state-of-the-art performance. The codes are available at https://github.com/Deriq-Qian-Dong/III-Retriever.

\end{abstract}




\maketitle
\section{Introduction}
\begin{figure}		
    \centering
    \subfloat[Cross-encoder]{\includegraphics[width=0.3\hsize,height=3cm]{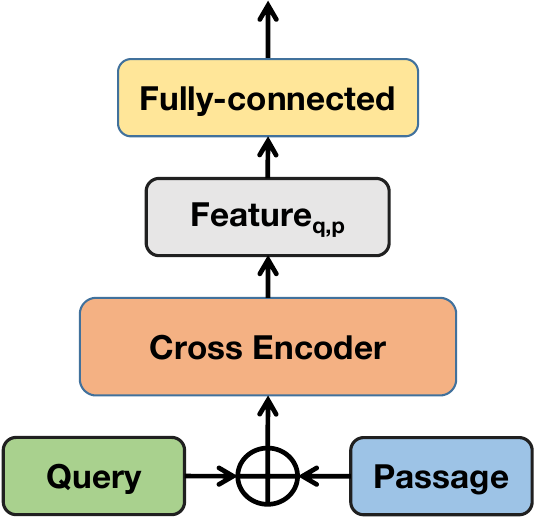}}\hspace{5mm}
    \subfloat[Dual-encoder]{\includegraphics[width=0.3\hsize,height=3cm]{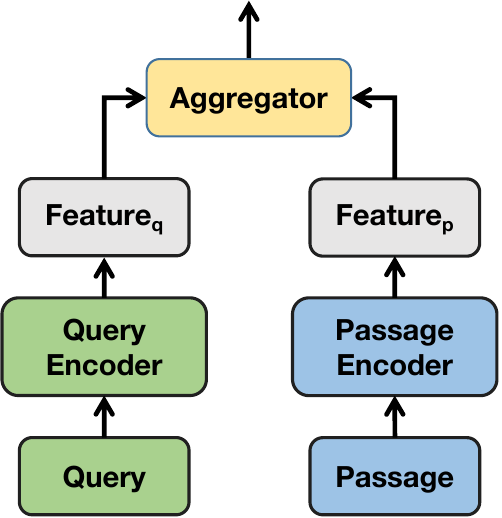}}
    
    \subfloat[Late-interaction encoder]{\includegraphics[width=0.4\hsize,height=4cm]{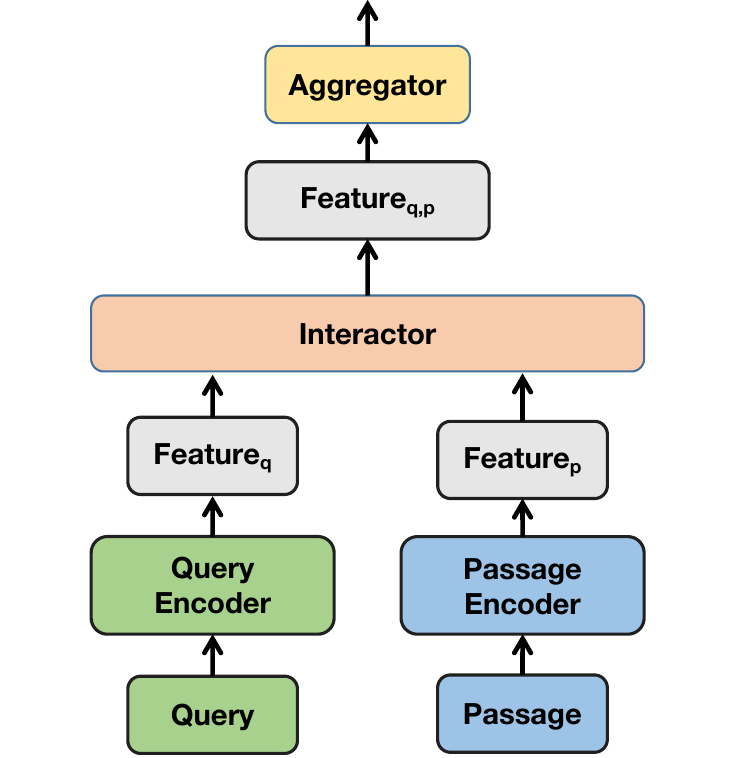}}\hspace{5mm}
    \subfloat[$\M$ retriever]{\includegraphics[width=0.5\hsize,height=4cm]{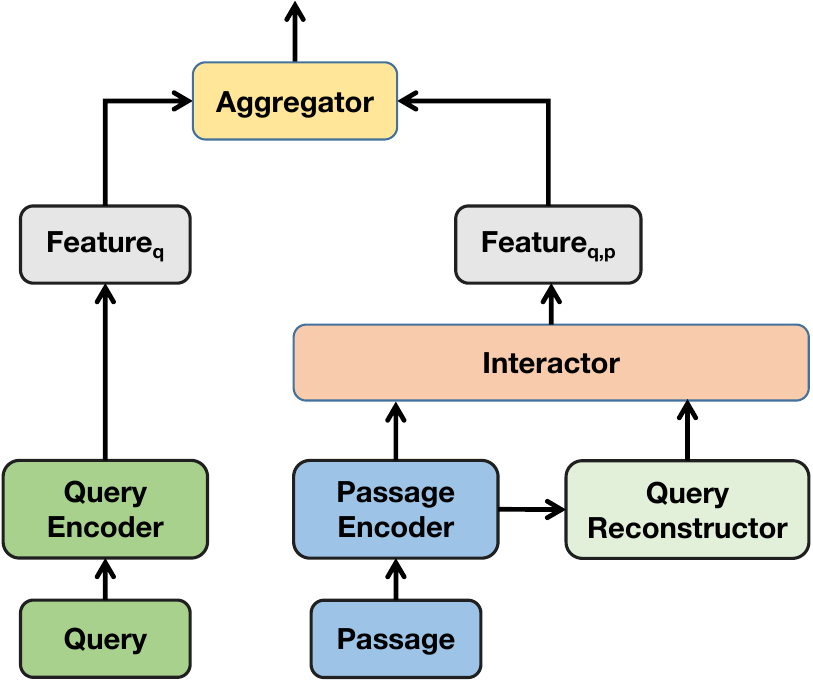}}
    \caption{Illustration of three conventional PLM-based IR models: (a) cross-encoder, (b) dual-encoder, (c) late-interaction encoder, and our proposed (d) $\M$ retriever.}
    \label{fig:interactionParadigms}
\end{figure}
Passage retrieval is 
fundamental
in modern information retrieval (IR) systems, typically serving as a preceding stage of reranking. 
The aim of passage retrieval is to find relevant passages 
from a large corpus for a given query, which
is crucial to the final ranking performance~\cite{karpukhin2020dense, xie2023t2ranking, cheng2023layout,li2023pretrained,zou2022pre}. Conventional methods for passage retrieval (e.g., BM25~\cite{robertson2009probabilistic}) usually consider lexical matching between the terms of query and passage. 
In recent years, neural retrievers based on pre-trained language models (PLMs) have prospered and achieved the state-of-the-art performance.


In particular, existing PLM-based IR models can be broadly categorized into cross-encoders~\cite{nogueira2019passage}, dual-encoders~\cite{karpukhin2020dense} and late-interaction encoders~\cite{gao2021coil,khattab2020colbert}, as shown in Figures~\ref{fig:interactionParadigms}(a), \ref{fig:interactionParadigms}(b) and \ref{fig:interactionParadigms}(c), respectively. 
Without considering the fine-grained interactions between the tokens of query and passage, the major merit of dual-encoders is their high efficiency in inference. Yet, their effectiveness is usually considered sub-optimal compared with cross-encoders or other interaction-based models.
Cross-encoders take the concatenation of query and passage as input to perform full interaction that effectively captures relevance features.
As query-passage interactions are important factors in relevance modeling~\cite{guo2016deep}, cross-encoders usually have superior ranking performance. However, their applications are limited to small collections (e.g., the top passages retrieved by dual-encoders) due to their high inference latency.
To combine the merits of both methods, late-interaction encoders adopt separate query/passage encoding and apply lightweight interaction schemes (i.e., late interactions) between the vectors of query and passage. 
They are usually more effective than dual-encoders for passage retrieval and less computationally expensive than cross-encoders.


Despite their effectiveness, late-interaction models are still sub-optimal for passage retrieval on large corpus, mainly due to two problems. First, effective late-interaction models usually relies on token-level representations of passages to allow subsequent token-level interactions~\cite{khattab2020colbert,santhanam2022colbertv2}, where the storage cost of such multi-vector passage representation is enormous.
Second, compared to dual-encoders, which adopts simple dot-product operation between single-vector representations of queries and passages,
late-interaction models still requires extra computation for each query-passage pair. Such cost could be magnified by the scale of massive corpus and eventually cause unacceptable efficiency degeneration~\cite{li2023constructing}. As such, existing late-interaction methods can hardly be applied to real-world scenarios that require low inference latency and storage cost.

To address these limitations and explore a 
better solution w.r.t. effectiveness and efficiency for passage retrieval, we propose a novel yet practical paradigm that $\textbf{I}$ncorporates $\textbf{I}$mplicit $\textbf{I}$nteraction ($\M$) in dual-encoders.
Unlike existing interaction schemes that requires explicit query text as input, the implicit interaction is conducted between a passage and 
the pseudo-query vectors generated from the passage.
Note that the generated pseudo-query vectors are implicit (i.e., latent) without explicit textual interpretation. 
Such implicit interaction paradigm is appealing, as 1) it is fully decoupled from actual query, and thus allows high online efficiency with offline caching of passage vectors, and 2) compared with using an off-the-shelf generative model~\cite{nogueira2019doc2query} to explicitly generate textual pseudo-query, our pseudo-query is represented by latent vectors that are jointly optimized with the dual-encoder backbone, which is more expressive for the downstream retrieval task.

To conduct implicit interaction, we propose a novel model architecture as shown in Figure \ref{fig:interactionParadigms}(d). It advances vanilla dual-encoders with two auxiliary modules.
First, we introduce a lightweight generative module namely \textbf{query reconstructor} to generate pseudo-query vectors for a given passage. 
Next, we apply a \textbf{query-passage interactor} that takes the concatenation of the passage vectors and the generated pseudo-query vectors as input, to perform implicit interaction. 
The interactor outputs query-aware passage vectors for each passage, which can be pre-computed and cached before deploying the model for online inference. 
The final query-passage relevance scores can be computed with simple dot-product operation,
which gives our model the same high efficiency and low storage cost as dual-encoders. 
The superior balance between effectiveness and efficiency makes our model more attractive in real-world applications. 
We summarize our main contributions as follows:

\begin{itemize}[leftmargin=5mm]
    \item 
    We propose a novel PLM-based retrieval model, namely $\M$ retriever, which incorporates implicit interaction in dual-encoders.
    \item We introduce two modules in $\M$ retriever that are jointly trained with query and passage encoders in an end-to-end manner, i.e., query reconstructor and query-passage interactor. The query reconstructor is able to generate pseudo-queries for the query-passage interactor, which subsequently encodes query-aware information in the final passage vectors.
    \item We conduct comprehensive evaluation on large scale datasets. The results show that $\M$ is able to achieve superior performance w.r.t both effectiveness and efficiency for passage retrieval.
    We also conduct a thorough study to clarify the effects of implicit interaction.
\end{itemize}
\section{Related Work}
In this section, we briefly review some existing studies with respect to three topics, i.e., traditional neural IR models, PLM-based IR models and query generation for IR.
\subsection{Conventional Neural IR Models}
Modern information retrieval systems usually adopt the two-stage paradigm, i.e. retrieval-then-reranking.
Neural IR models can be categorized as either retrievers or rerankers based on their served stage. 
Retrievers can pre-compute the vector representation of passages in corpus and thus perform efficient retrieval via approximate nearest neighbor algorithms. 
Therefore, retrievers usually define sophisticated representation learning module. 
DSSM~\cite{huang2013learning} is a representative neural retriever, which uses the fully-connected network for representation learning.
Besides, convolutional networks~\cite{hu2014convolutional,qiu2015convolutional,shen2014latent, dong2021legal} and recurrent networks~\cite{palangi2016deep,wan2016deep} are also widely used for representation learning in neural retrievers.
On the other hand, rerankers effectively capture relevance features through sufficient interactions.
The interaction module plays a vital role in the effectiveness of rerankers.
DRMM~\cite{guo2016deep} designs a matching histogram mapping to model the interaction between terms of query and passage.
Conv-KNRM~\cite{dai2018convolutional} and Arc-II~\cite{hu2014convolutional} use convolutional networks as the interaction module.
However, the computational overhead of rerankers during inference is significantly higher than that of retrievers, and thus rerankers only serve a small set of candidates at the final stage.
\subsection{PLM-based IR models}
\noindent \textbf{PLM-based retriever.}
PLM-based retrievers usually compute low dimensional representations for the query and passage using the encoder of pre-trained transformer~\cite{vaswani2017attention}, such as BERT~\cite{devlin2018bert} and RoBERTa~\cite{liu2019roberta}. 
DPR~\cite{karpukhin2020dense} is the first to leverage PLM for the task of semantic retrieval, while extensive methods are subsequently proposed to improve the effectiveness.
In particular, most of the existing PLM-based retrievers 
improve the model performance from the following aspects.
(1) \textit{\textbf{By introducing late-interactions after encoding:}} ColBERT~\cite{khattab2020colbert}, COIL~\cite{gao2021coil} and ME-BERT~\cite{luan2021sparse} are three representative studies that explicitly model the interactions after query/passage encodings. The performance is largely boosted by the interactions compared with DPR (i.e., vanilla dual-encoder), while the late interaction also brings significant computational overhead. 
(2) \textit{\textbf{By designing effective fine-tuning processes:}}
For example,  ANCE~\cite{xiong2020approximate} proposes to a hard negative sampling technique that greatly improve the effectiveness.
Moreover, RocketQAv1~\cite{qu2021rocketqa} and RocketQAv2~\cite{ren2021rocketqav2} 
boost the performance of dense retrieval models by leveraging the power of cross-encoder. 
The relevance features captured through sufficient interaction by the cross-encoder could be properly transferred to retrievers in a cascade or joint training manner.
ERNIE-Search~\cite{lu2022ernie} narrows the divide between cross-encoder and dual-encoder models through on-the-fly distillation in the process of fine-tuning. ColBERTv2~\cite{santhanam2022colbertv2} further improves ColBERT by employing fine-tuning with distillation.
(3) \textit{\textbf{By designing pre-training tasks tailored for retrieval:}} A handful of studies focused on constructing pseudo-training data for retrieval-oriented pre-training, such as ICT~\cite{chang2020pre}, COSTA~\cite{ma2022pre}, DCE~\cite{li2022learning}, etc.
Besides, several studies~\cite{lu2021less, zhou2022master, wu2022contextual, wu2022query, wu2022contextual, wang2022simlm, liu2022retromae, li2023sailer} employ weak generative modules (i.e. decoder)
to enhance the query/passage encoding 
through pre-training. 
Notably, after pre-training, the weak decoder is discarded and only the enhanced encoder is employed as the backbone of retriever.
In this work, we propose a novel approach that incorporates implicit interaction modeling into the dual-encoder architecture by introducing a generative module. 
To the best of our knowledge, this is the first attempt to introduce a generative module as a backbone in a retriever.

\noindent \textbf{PLM-based reranker.}
PLM-based rerankers usually take
the concatenated query and passage as input and perform full interaction between query and passage via self-attention~\cite{guo2020deep,dong2021latent, dong2022disentangled}.
In particular, monoBERT~\cite{nogueira2019passage} is the first work that re-purpose BERT as a reranker. duoBERT~\cite{nogueira2019multi} integrates monoBERT in a multistage ranking pipeline and further adopts a pairwise classification framework for the final re-ranking.
UED~\cite{yan2021unified} utilizes a unified encoder-decoder framework to jointly optimize passage reranking and query generation tasks, demonstrating that these two tasks could facilitate each other.
KERM~\cite{dong2022incorporating} leverages external knowledge graph to more accurately model the interaction between query and passage, and thus achieves the state-of-the-art results. Inspired by the superior performance of PLM-based reranker, our method is equipped with a cross-interaction module that allows effective implicit interaction during passage encoding.

\subsection{Query Generation for IR}
The technique of query generation
has been widely adopted in a variety of IR applications. For example, 
a well-known query generation method, namely doc2query~\cite{nogueira2019document}, proposes a sequence-to-sequence model trained on relevant query-passage pairs to generate multiple queries for each passage. These generated queries can be considered as a passage expansion for the downstream retrieval task.
This approach is effective in mitigating the issue of term mismatch between queries and passages. 
Moreover, docT5query~\cite{nogueira2019doc2query} employs T5~\cite{raffel2020exploring} to generate queries and delivers an improved performance over doc2query.
More recently, the application of query generation has been examined in the context of pre-training dense retrievers~\cite{wu2022query}, data augmentation~\cite{liang2020embedding,li2022learning,bonifacio2022inpars} and domain adaptation~\cite{ma2020zero, wang2021gpl, dai2022promptagator, xiao2023social4rec}. 
However, these studies leverage query generation models as an off-the-shelf tool, which might not be the optimal for the downstream retrieval task.
In our study, we introduce a lightweight generative module, i.e., the query reconstructor, which is jointly trained with the retrieval backbone in an end-to-end manner. By doing this, the query reconstructor is learned to generate pseudo-queries that are more helpful for the final retrieval task.

\section{PRELIMINARIES}
In this section, we introduce the problem definition of passage retrieval,
and present several PLM-based IR methods.
\subsection{Problem Definition}
Modern IR systems usually follow a retrieve-then-rerank pipeline.
Given a corpus of passages $\mathcal{G}=\{\mathbf{p}_i\}_{i=1}^{G}$ 
, the aim of \textbf{retrieval} is to find a small set of candidate passages (i.e., $\mathcal{K}=\{\mathbf{p}_j^\mathbf{q}\}_{j=1}^{K}$) and $K \ll G$) that is relevant to 
a specific query $\mathbf{q}$.
In particular, a passage $\mathbf{p}$ is a sequence of words $\mathbf{p}=\{w_p\}_{p=1}^{|\mathbf{p}|}$, where $|\mathbf{p}|$ denotes the length of $\mathbf{p}$. Similarly, a query is a sequence of words $\mathbf{q}=\{w_q\}_{q=1}^{|\mathbf{q}|}$. 
After the retrieval stage, reranking is conducted to finalize a better permutation on $\mathcal{K}$, where more relevant passages are ranked higher.

It worth noting that retrieval and reranking models usually have different practical concerns. In particular, both efficiency and effectiveness are vital for retrieval models, as real-world scenarios usually require fast retrieval on large scale corpus. On the other hand, reranking models are more concentrated on effectiveness, and they 
should be able to effectively capture the subtle differences between relevant passages.
In this work, our attention is focused on the PLM-based retriever, and we propose a implicit interaction paradigm that achieves the state-of-the-art performance in terms of both effectiveness and efficiency for passage retrieval. 


\subsection{PLM-based Retriever and Reranker}
The performance of neural IR models, including retrievers and rerankers, have been significantly boosted by pre-trained language models (PLM), where various ways of leveraging PLM for IR are proposed.
As illustrated in Figure~\ref{fig:interactionParadigms}, PLM-based IR models can be categorized into three types, i.e., dual-encoders, late-interaction encoders and cross-encoders, in terms of the interaction mechanism applied between query and passage. Overall, existing studies indicate that incorporating more interactions between queries and passages in a PLM-based IR method can improve 
relevance modeling, but it also comes at the cost of extra computational overhead.
In the following, we further introduce the detailed structures of these models.

\noindent \textbf{Dual-encoder.} Dual-encoders employ two PLM-based encoders to respectively encode the query and passage in a latent embedding space. The relevance score $S{(\mathbf{q},\mathbf{p})}$ between query and passage is formulated as
\begin{equation}
    \label{eq:dualencoder}
    S{(\mathbf{q},\mathbf{p})}=\operatorname{Aggregate}\left(\mathbb{E}_q(\mathbf{q})_{[CLS]},\mathbb{E}_p(\mathbf{p})_{[CLS]}\right).
\end{equation}
Here, $\operatorname{Aggregate}(\cdot)$ is usually implemented as a simple metric (e.g., dot-product) between query and passage vectors, which is computed by query and passage encoders (i.e., $\mathbb{E}_q$ and $\mathbb{E}_p$), respectively. The encoders are stacked transformer layers, where we fetch the 
representation of [CLS] token in the last layer as
final query/passage vector. 

The major merit of dual-encoders lies in its high efficiency. As the query and passage are decoupled at encoding, the passages in large corpus $\mathcal{G}$ can be pre-computed and cached offline. By doing this, substantial computational resources could be saved during the online inference for fast retrieval. However, the limitation is also apparent. 
The absence of interaction between the query and passage during their encoding
leads to
an inability to effectively capture complex relevance~\cite{humeau2019poly,khattab2020colbert,ye2022fast}.

\noindent \textbf{Cross-encoder.} Cross-encoders are considered the most effective 
PLM-based IR method due to their early incorporation of query-passage interactions. It takes the concatenation of query and passage as input, and computes the relevance score as
\begin{equation}
    \label{eq:crossencoder}
    S{(\mathbf{q},\mathbf{p})}=\operatorname{FC}\left(\mathbb{E}_{q,p}(\mathbf{q}\oplus\mathbf{p})_{[CLS]}\right),
\end{equation}
where $\oplus$ means the concatenation operation and $\mathbb{E}_{q,p}$ is the PLM encoder. 
The $\operatorname{FC}$ is a fully-connected layer that transforms the [CLS] representation to a relevance score.

Cross-encoders allow full token-level interactions between query and passage via self-attention~\cite{devlin2018bert}, where 
relevance features could be adequately captured.
This leads to a superior performance in relevance modeling compared with other PLM-based IR models.
However, compared with dual-encoders, cross-encoders require extensive online computation, where no intermediate representations could be pre-computed and cached offline.
The low efficiency of cross-encoders limits its application for retrieval on large scale corpus, and thus they are mainly designed for reranking stage.

\noindent \textbf{Late-interaction encoder.} To balance efficiency and effectiveness, the late-interaction paradigm introduces interaction between query and passage after encoding, which can be formulated as 
\begin{equation}
    \label{eq:lateencoder}
    S{(\mathbf{q},\mathbf{p})}=\operatorname{Aggregate}\left(\operatorname{Interact}\left(\mathbb{E}_q(\mathbf{q}), \mathbb{E}_p(\mathbf{p})\right)\right).
\end{equation}
The $\operatorname{Aggregate}(\cdot)$ operation aggregates the relevance features captured from the $\operatorname{Interact}(\cdot)$ into a relevance score $S{(\mathbf{q},\mathbf{p})}$. 

ColBERT~\cite{khattab2020colbert} is a representative of late-interaction method.
Its interaction is implemented as the
maximum similarity score between each pair of token representations of query and passage in the final layers.
Then, these scores are aggregated into a final relevance score, which can be formulated as
\begin{equation}
    \label{eq:colbert}
    S{(\mathbf{q},\mathbf{p})}=\sum_{q=1}^{|\mathbf{q}|}\max_{p=1}^{\mathbf{|\mathbf{p}|}}\left(\mathbb{E}_q(\mathbf{q})_{w_q} \cdot \mathbb{E}_p(\mathbf{p})_{w_p}\right).
\end{equation}
To reduce the computational overhead of ColBERT, COIL~\cite{gao2021coil} restricts the interactions to occur solely between pairs of query and passage tokens that have an exact match. 
More details about these methods can be found in their original papers~\cite{gao2021coil,khattab2020colbert}.

Similar to dual-encoders, late-interaction encoders also decouple the encoding of query and passage, and thus
allow pre-computation of all passage vectors in corpus $\mathcal{G}$. 
However, the late interactions still create considerable computational overhead for each query-passage pair. Worse still, they further cost enormous space footprint for caching multi-vector passage vectors, where dual-encoders only need to store single-vector passage vectors.


\noindent\textbf{Remarks.} Overall, former experience tells us that effective interaction usually cost extra computation or storage, where most of the existing studies are proposed to make amends. However, we intend to investigate a different research question:
\textbf{Can we model query-passage interaction without any efficiency degeneration?}
Noting that the dual-encoders are efficient due to its offline pre-computation of passage vectors, the key to answer this question is how to pre-compute and model query-passage interactions offline. However, this is challenging because the actual queries issued by users are agnostic during the pre-computation, while we can only access to the passages in the corpus. In the next section, we propose a novel method, namely $\M$ retriever, which tackles this challenge to achieve high effectiveness without hurting efficiency.


\section{Method}
In this section, we present $\M$ retriever, an effective approach that incorporates implicit interaction in dual-encoder. 
We first introduce the overall architecture, which includes query and passage encoders, query reconstructor and query-passage interactor. 
Then, we present the details of implicit interaction, and the end-to-end optimization and inference of $\M$ retriever.
\begin{figure}
		\centering
		\includegraphics[width=\linewidth]{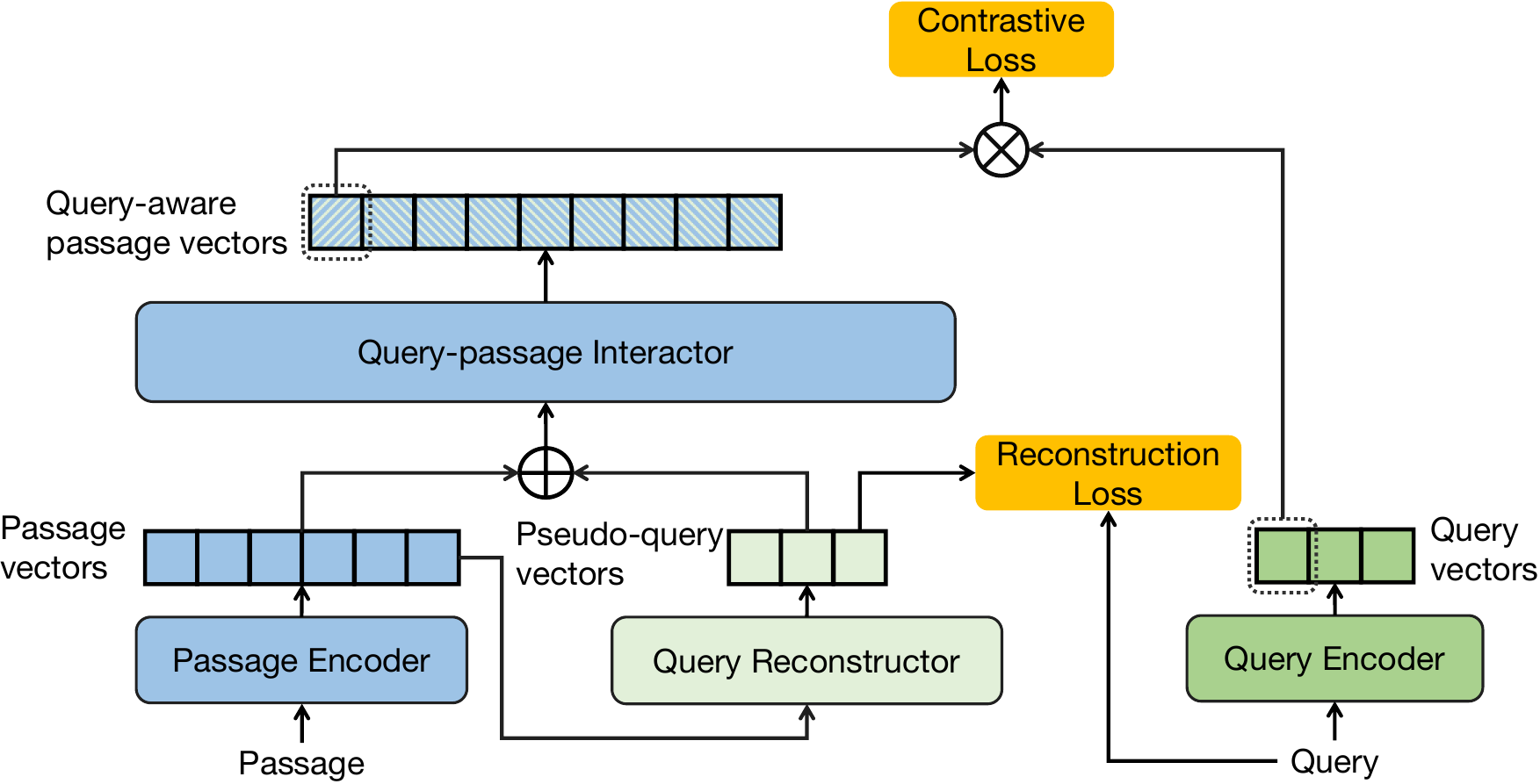}
		\caption{The architecture of $\M$ retriever.}
		\label{fig:architecture}
\end{figure}
\subsection{Overall Architecture}
\label{sec:architecture}
Figure~\ref{fig:architecture} illustrates the overall architecture of the $\M$ retriever. In particular, we advance the passage encoder of vanilla dual-encoders with two auxiliary modules, i.e., query reconstructor and query-passage interactor. Overall, the workflow of $\M$ can be formulated as follows:
\begin{itemize}[leftmargin=5mm]
    \item \textbf{Query/Passage encoding}. The query and passage encoders (i.e., vanilla dual-encoders) are the backbone of our proposed method. They first encode the tokens of query and passage into latent vectors.
    \item \textbf{Query reconstruction}. Inspired by generative models~\cite{radford2018improving}, we introduce a lightweight query reconstructor to generate a pseudo-query for each passage, which can be viewed as a potential query for a specific passage. 
    \item \textbf{Query-passage interaction}. We apply a query-passage interactor to conduct cross-encoder-alike interaction between each passage and its pseudo-query.
    It finalizes a \textit{query-aware passage vector}, which learns to encode passage information that are vital to its potential query. 
    \item \textbf{Relevance computation}. The final relevance score is computed as the dot-product between the query vector produced by query encoder, and the query-aware passage vector produced by the query-passage interactor. The simple relevance metric allows high efficiency for online retrieval.
\end{itemize}
We refer to such interaction over vanilla dual-encoders as \textbf{implicit interaction}, since it solely relies on generated pseudo-query vectors, rather than textual query terms. Note that the inference of implicit interaction is conducted on the passage side, and thus it could be pre-computed and cached to enable efficient online retrieval. Next, we focus on the two auxiliary modules and  elaborate how they are incorporated to conduct implicit interaction.

\subsection{Incorporating Implicit Interaction}

\noindent \textbf{Query reconstructor.} 
The query reconstructor is a generative model with stacked transformer layers, which can be viewed as a decoder module for passage encoder. In particular, it takes a set of trainable embedding $\mathbf{I}^0\in\mathbb{R}^{\overline{q}\times d_{model}}$ as input vectors, and conduct cross-attention with the output vectors of passage encoding. For simplicity, we use the special token, [MASK], as the initial parameters of $\mathbf{I}^0$.
Here, $\overline{q}$ is the length of generated queries and $d_{model}$ is the dimension of the embeddings. 
In each layer $n=1,...,N$, the output vectors $\mathbf{I}^n$ are computed as
\begin{equation}
    \mathcal{A}_{(\mathbf{I}^{n-1}, p)}=\text{softmax}(\frac{(\mathbf{W}_n^Q\mathbf{I}^{n-1})(\mathbf{W}_n^K\mathbb{E}_p(\mathbf{p}))^{\textrm{T}}}{\sqrt{d_{model}}}),
    \label{eq:attenmatrix}
\end{equation}
\begin{equation}
    \mathbf{I}^n = \sum_{p=1}^{|\mathbf{p}|} \mathcal{A}_{(\mathbf{I}^{n-1}, p)}\mathbf{W}_n^V\mathbb{E}_p(\mathbf{p}),
\end{equation}
where $\mathcal{A}_{(\mathbf{I}^{n-1}, p)}$ is the cross-attention between $\mathbf{I}^{n-1}$ and the passage vectors $\mathbb{E}_p(\mathbf{p})$, and $\mathbf{W}_n^{*}$ are the parameters of query reconstructor.
The reconstructed pseudo-query vectors for passage $\mathbf{p}$ is denoted as $\mathbb{K}_q(\mathbf{p})\coloneqq\mathbf{I}^{N}$. 
Notably, the input embedding $\mathbf{I}^0$ is the same for all passages. By doing this, we can reconstruct pseudo-query vectors $\mathbb{K}_q(\mathbf{p})$ from passage vectors in a query agnostic manner.

It worth mentioning that the query reconstructor differs from existing generative language models from two perspectives: 1) Unlike conventional auto-regressive models that generate tokens sequentially, our query reconstructor generates all the $\overline{q}$ vectors in parallel, and thus is more efficient;
2) Our model generates latent vectors rather than actual words to represent the pseudo-query, which is more expressive to represent semantic information for downstream retrieval task.

\noindent \textbf{Query-passage interactor.} The interactor $\mathbb{E}_{q,p}(\cdot)$ has a cross-encoder-alike structure that stacks multiple transformer layers. It conducts full cross-interaction between passage vectors $\mathbb{E}_p(\mathbf{p})$ and its reconstructed pseudo-query vectors $\mathbb{K}_q(\mathbf{p})$, i.e., implicit interaction.
The interactor refines the passage vectors and outputs query-aware passage vectors. Intuitively, the interactor leverages the pseudo-query to encode important knowledge in the query-aware passage vector that might be relevant to real queries.
More formally, the  $\mathbb{T}_p(\mathbf{p})$ are computed as 
\begin{equation}
    \mathbb{T}_p(\mathbf{p})=\mathbb{E}_{q,p}\left(\mathbb{K}_q(\mathbf{p})\oplus\mathbb{E}_p(\mathbf{p})\right),
\end{equation}
where $\oplus$ means the concatenation operation. 
Finally, we can advance vanilla dual-encoders by rewriting the relevance score $S{(\mathbf{q},\mathbf{p})}$ in Eq.~\ref{eq:dualencoder} as
\begin{equation}
S{(\mathbf{q},\mathbf{p})}=\mathbb{E}_q(\mathbf{q})_{[CLS]}\cdot\mathbb{T}_p(\mathbf{p})_{[CLS]}.
\end{equation}

By introducing query reconstructor and query-passage interactor in passage encoding, our $\M$ retriever is effective and efficient, as 1) it effectively incorporates implicit interaction that encodes vital passage information w.r.t. potential queries, and 2) the implicit interaction is conducted on the passage side in a query agnostic manner, which brings high online inference efficiency that is on par with the vanilla dual-encoder.

\subsection{Model Optimization}
\label{sec:optimization}
\begin{figure}
		\centering
		\includegraphics[width=0.75\linewidth]{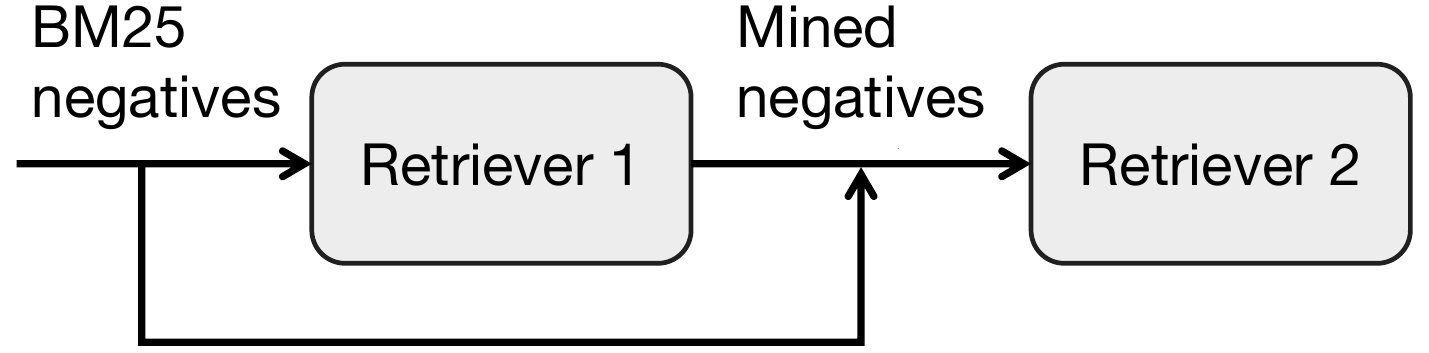}
		\caption{Illustration of our fine-tuning pipeline.}
		\label{fig:finetune}
\end{figure}

\noindent\textbf{Retrieval loss}. Following previous work~\cite{gao2021unsupervised}, our $\M$ retriever is optimized by the following contrastive loss
\begin{equation}
\label{eq:contrastive}
\mathcal{L}_{c}=-\log \frac{\exp \left(S\left(\mathbf{q}, \mathbf{p}_{+}\right)\right)}{\exp \left(S\left(\mathbf{q}, \mathbf{p}_{+}\right)\right)+\sum\limits_{\mathbf{p}_{-}\in \mathcal{N}_{-}} \exp \left(S\left(\mathbf{q}, \mathbf{p}_{-}\right)\right)},
\end{equation}
where $\mathcal{N}_{-}$ is a set of hard negative passages (denoted as $\mathbf{p}_{-}$) for query $\mathbf{q}$.
As illustrated in Figure~\ref{fig:finetune}, the fine-tuning process consists of two stages, where the optimized models are called retriever 1 and retriever 2, respectively.
During the training of retriever 1, the negative samples $\mathcal{N}_{-}$ are BM25 hard negatives.
During the training of retriever 2, hard negatives are also mined using the optimized retriever 1 to complement the negative pool $\mathcal{N}_{-}$.

\noindent\textbf{Reconstruction loss}. In addition to the retrieval loss (i.e., Eq. \ref{eq:contrastive}), we also introduce an auxiliary reconstruction loss to guide the query reconstructor, 
which is defined as
\begin{equation}
\label{eq:reconstructor}
    \mathcal{L}_r = 
    \sum\limits_{w_i\in \mathbf{q}}\operatorname{CrossEntropy}\left(\mathbf{W}^R\mathbb{K}_q(\mathbf{p})_i, w_i\right),
\end{equation}
where $w_i$ is the $i$-th word of a pseudo query $\mathbf{q}$.
The pseudo query could be generated by RACE~\cite{rose2010automatic} or other keyword extraction methods, such as large language models.
$\mathbf{W}^R$ is the parameter of reconstructor, mapping the dimension of $\mathbb{K}_q(\mathbf{p})_i$ from $d_{model}$ to vocabulary size.

The final training loss of $\M$ retriever is the combination of the above-mentioned two losses as
\begin{equation}
\label{eq:loss}
    \mathcal{L} = \mathcal{L}_c + \lambda\mathcal{L}_r,
\end{equation}
where $\lambda$ is a hyper-parameter. All the modules are jointly optimized with this loss in an end-to-end manner.
\subsection{Model Inference}
\noindent\textbf{Offline pre-computation}. 
The computation of query-aware passage vectors $\mathbb{T}_p(\mathbf{p})$ is totally decoupled with online inference w.r.t. a specific query. Therefore, all the passage vectors in the corpus $\mathcal{G}$ could be pre-computed and stored. Note that the pre-computed passage vectors are interacted with pseudo-query vectors generated by the query reconstructor, and thus are more expressive than the passage vectors produced by vanilla dual-encoders. 
Besides, we adopt single-vector representation for each passage, which avoids massive storage cost.

\noindent\textbf{Online inference}. 
The online inference process is identical to vanilla dual-encoders, and thus our method has the same high efficiency. When a query is received, 
$\M$ applies the query encoder to compute its vector $\mathbb{E}_q(\mathbf{q})_{[CLS]}$. Next, it conducts maximum inner product search (MIPS)
over the offline-cached query-aware passage vectors $\{\mathbb{T}_p(\mathbf{p}_i)_{[CLS]}\}_{i=1}^{G}$ to retrieve a set of relevant passages.

\noindent\textbf{Analysis}. For the online inference stage, the time complexity of $\M$ retriever is $\mathrm{O}\left(E+G\right)$, where $E$ and $G$ are the cost of query encoding and MIPS operation over the corpus $\mathcal{G}$, respectively.
For late-interaction encoder with multi-vector representations, their time complexity is $\mathrm{O}\left(E+Q\times P\times G\right)$, where $Q$ and $P$ indicate the number of vectors representing query and passage, respectively. Moreover, our single-vector representation could be more easily supported by commonly-used indexing techniques~\cite{johnson2019billion,malkov2018efficient}
Therefore, our method has superior efficiency compared with existing late-interaction encoders.

\section{Experimental Setup}
\begin{table}[t]
	\centering
	\caption{Statistics of MSMARCO-DEV and TREC DL 19.}
	\label{tab:queries}
	\begin{tabular}{l|c|c}
		\hline
        \hline
        & MSMARCO-DEV & TREC DL 19 \\
		\hline
		\#Queries   & 6980        & 43\\
		\hline
		\#Rel.Psgs.  & 7437          & 4102\\
        \hline
        Rel.Psgs./Query & 1.1 & 95.4 \\
        \hline
        \#Graded.Labels &2 & 4\\
		\hline
        \hline
	\end{tabular}
\end{table}
\subsection{Datasets}
We use MSMARCO-Passage~\cite{nguyen2016ms} as the large-scale corpus for our experiments. It
consists of around 8.8 million passages.
Following previous
work~\cite{karpukhin2020dense,xiong2020approximate,gao2021coil,khattab2020colbert}, we train our model on MSMARCO-TRAIN query set including 502,939 queries, and evaluated on two widely used query sets, i.e., MSMARCO-DEV and TREC DL 19. 
\textbf{MSMARCO-DEV}~\cite{nguyen2016ms} includes 6,980 sparsely-judged queries,
each of which 
has 1.1 relevant passages on average.
\textbf{TREC DL 19}~\cite{craswell2020overview} 
contains 43 densely-judged queries, which are annotated with fine-grained relevance labels, i.e., irrelevant, relevant, highly relevant and perfectly relevant. Such data can be used to evaluate fine-grained ranking performance.
Tabel~\ref{tab:queries} summarizes the detailed information of the two query sets. 

\subsection{Baselines}
We include the following variants of our methods to ensure a fair comparison with baselines:
\begin{itemize}
    \item \textbf{$\M$ retriever$_1$} is our proposed method that incorporates implicit interaction in dense retrieval.
    \item \textbf{$\M$ retriever$_2$} is an improved version of $\M$ retriever$_1$, which further leverages the widely-used negative sampling technique~\cite{xiong2020approximate}.
    \item \textbf{$\M$ retriever$_3$} is initialized from RetroMAE~\cite{liu2022retromae} and fine-tuned with hard negatives, following the baselines.
    \item \textbf{$\M$ retriever$_4$} is also initialized from RetroMAE~\cite{liu2022retromae} and further distilled using a cross-encoder with the Kullback–Leibler divergence loss function.
\end{itemize}

\textbf{$\M$ retriever$_1$} and \textbf{$\M$ retriever$_2$} are compared with dense methods without special pre-training and distillation.
We include two sparse retrievers, i.e., BM25~\cite{robertson2009probabilistic} and DeepCT~\cite{dai2020context}, 
as baselines.
We include more dense retrievers, which can be categorized as 
non-interaction, late-interaction, and early-interaction methods.
1) \textbf{Non-interaction methods:} DPR~\cite{karpukhin2020dense} and ANCE~\cite{xiong2020approximate} are two widely used baselines that do not consider any form of query-passage interaction; 2) \textbf{Late-interaction methods:} ME-BERT~\cite{luan2021sparse}, COIL~\cite{gao2021coil} and ColBER~\cite{khattab2020colbert} apply lightweight interaction after query/passage encoding; 3) \textbf{Early-interaction methods:} DRPQ~\cite{tang2021improving} and DCE~\cite{li2022learning} model the interaction during the encoding stage. 
Notably, both DCE and our proposed $\M$ retriever conduct interaction between pseudo-query and passage during passage encoding. The key difference lies in that DCE employs docT5Query~\cite{nogueira2019doc2query} to explicitly generate pseudo-queries, while $\M$ retriever utilizes a lightweight reconstruction module to implicitly reconstruct pseudo-query vectors in an end-to-end manner.

\textbf{$\M$ retriever$_3$} and \textbf{$\M$ retriever$_4$} are compared with baselines with special pre-training and distillation, respectively.
For dense retrieval models with task-specific pre-training, we include the following methods:
coCondenser~\cite{gao2021unsupervised} continues to pre-trained on the target corpus with contrastive loss. Other pre-trained methods, such as SimLM~\cite{wang2022simlm}, Cot-MAE~\cite{wu2022contextual} and RetroMAE~\cite{liu2022retromae}, employ a bottleneck architecture that learns to compress the passage information into a vector through pre-training.
We also include the state-of-the-art methods that facilitate dense retrieval with knowledge distillation:	
TAS-B~\cite{hofstatter2021efficiently}, RocketQAv2~\cite{ren2021rocketqav2} and ERNIE-Search~\cite{lu2022ernie} primarily concentrate on distilling knowledge from a cross-encoder to a single vector retriever. On the other hand, SPLADEv2~\cite{formal2021splade} and ColBERTv2~\cite{santhanam2022colbertv2} focus on distilling knowledge from a cross-encoder to a multi-vector retriever. All baselines with special pre-training or distillation can be categorized as non-interaction retrievers, except for SPLADEv2~\cite{formal2021splade} and ColBERTv2~\cite{santhanam2022colbertv2}.

\begin{table*}[!htp]
\centering
\caption{Performance comparison on MARCO-DEV and TREC DL 19.}
\label{tab:performance}
\begin{threeparttable}
\begin{tabular}{l|c|c|c|cc|c}
\hline 
\hline
\multirow{2}{*}{ Method }
&\multicolumn{3}{c|} { Settings }& \multicolumn{2}{c|} { MARCO DEV Passage } & \multicolumn{1}{c} { TREC DL 19}\\
\cline{2-7} & Single vector? & Mined-negatives &Interaction&MRR@10&Recall@1000& NDCG@10 \\
\hline 
BM25 (anserini)        & - & - & - & .187 & .857     & .501 \\
DeepCT &-&-& - & $.243$ & $.905$& $.551$\\
\hline
\multicolumn{7}{c} { \textit{Comparison with dense methods without pre-training or distillation} } \\
\hline
DPR &\ding{51} & &Non-interaction&.314&.953&.590 \\
ANCE &\ding{51} & \ding{51} &Non-interaction&.330&.959&.648 \\
DCE & & &Explicit-early& .338 & -& - \\
DRPQ & & \ding{51} &Explicit-early&.345&.964&- \\
ME-BERT & &\ding{51}&Explicit-late&.334&-&.687 \\
COIL & & &Explicit-late&.355&.963&.704 \\
ColBERT & & &Explicit-late&.360&.968&.694\\
$\M$ retriever$_{1}$ & \ding{51} & &Implicit-early& .349 & .966 &.720\\
$\M$ retriever$_{2}$ & \ding{51} &\ding{51}&Implicit-early& $\textbf{.366}$ & $\textbf{.976}$& $\textbf{.727}$ \\
\hline 
\multicolumn{7}{c} { \textit{Comparison with dense methods with special pre-training} } \\
\hline
coCondenser & \ding{51} &\ding{51} & Non-interaction & .382 &.984 & .684 \\
SimLM & \ding{51} &\ding{51} &Non-interaction&.391&.986&- \\
Cot-MAE & \ding{51} &\ding{51} &Non-interaction&.394&$\textbf{.987}$&- \\
RetroMAE & \ding{51} &\ding{51} &Non-interaction&.393&.985&- \\
$\M$ retriever$_{3}$ & \ding{51} &\ding{51}&Implicit-early& $\textbf{.403}$ & $\textbf{.987}$ & $\textbf{.729}$ \\

\hline
\multicolumn{7}{c} { \textit{Comparison with dense methods with distillation} } \\
\hline
TAS-B  & \ding{51} & \ding{51}& Non-interaction &.340&.975&.712 \\
SPLADEv2 & & \ding{51}&Explicit-late&.368&.979&.729 \\
RocketQAv2 & \ding{51} & \ding{51}& Non-interaction &.388&.981&- \\
ColBERTv2 & & \ding{51}&Explicit-late&.397&.984&- \\
ERNIE-Search & \ding{51} & \ding{51}&Non-interaction&.401&.982&- \\
SimLM & \ding{51} & \ding{51}&Non-interaction&.411&.987&.714 \\
Cot-MAE  & \ding{51} &\ding{51} &Non-interaction&.404&.987&- \\
RetroMAE & \ding{51} & \ding{51}&Non-interaction&.416&$\textbf{.988}$&.681 \\
$\M$ retriever$_{4}$ & \ding{51} &\ding{51}&Implicit-early& $\textbf{.418}$ & $\textbf{.988}$ & $\textbf{.731}$ \\
\hline
\hline
\end{tabular}
   \end{threeparttable}
\end{table*}

\subsection{Implementation Details}
\label{sec:ImplementationDetails}
For training $\M$, we use the Lamb optimizer~\cite{you2019large} with a learning rate of 2e-5. The model is trained with a batch size of 16. The ratio of positive and hard negatives is set to 1:127 in the contrastive loss (i.e., Eq.~\ref{eq:contrastive}). Besides, the hyper-parameter $\lambda$ in Eq.~\ref{eq:loss} is decayed with epochs exponentially, starting from an initial value of 1.

For the comparison with dense methods without distillation or special pre-training, all the baselines are initialized with $\textrm{BERT}_{\textrm{base}}$ model, except ANCE~\cite{xiong2020approximate}, which utilizes $\textrm{RoBERTa}_{\textrm{base}}$. 
In our model, we set the number of layers of query encoder, passage encoder, query reconstructor and query-passage interactor as 6, 6, 3 and 3, respectively. We configure the length of generated query $\overline{q}$ as 32 to cover the majority of queries in the training data.
As such, $\M$ retriever has a comparable model size with the baselines on the passage side (i.e., 6+3+3 transformer layers), but fewer parameters on the query side.
The query and passage encoders are initialized
with $\textrm{BERT}_{\textrm{distill}}$. 
For the comparison with distillation or special pre-training, we directly use the RetroMAE~\cite{liu2022retromae} to initialize the backbone of $\M$, as pre-training is not the main focus of this paper. To minimize the number of parameters introduced, we configure the query reconstructor and query-passage interactor to consist of a single layer.	
The query reconstructor and query-passage interactor are random initialized.
Prior to fine-tuning, the query reconstructor and query-passage interactor undergoes optimization for 20K steps on passage collection $\mathcal{G}$ via Eq.~\ref{eq:loss} with $\lambda=1$ while keeping the parameters of backbone frozen. 
The pseudo query, generated by a language model Flan-T5-XL~\cite{chung2022scaling} in a zero shot setting, along with its corresponding passage, is regarded as a positive pair.

Our proposed model is implemented with PyTorch and Huggingface~\footnote{https://github.com/huggingface/transformers}. 
All the training and evaluation are conducted on 8 NVIDIA Tesla A100 GPUs (with 40G RAM). 

\section{EXPERIMENTAL RESULTS}

In this section, we present the experimental results and conduct thorough analysis of $\M$ to clarify its advantages.
\subsection{Overall Comparison}

\noindent \textbf{Effectiveness.} We first compare the effectiveness of $\M$ with all the baselines. The results are shown in Table~\ref{tab:performance},
where the detailed setting of each method is also included, i.e., whether a method employs single vector passage representation, negative mining or a particular interaction scheme.
Notably, the baselines are categorized into three groups: methods without special pre-training or distillation, methods with special pre-training, and methods with distillation. 
We report MRR@10 and Recall@100 on MARCO DEV Passage, and NDCG@10 on TREC DL 19.

First, we can draw several key findings from the first group (i.e., methods without pre-training or distillation):
\begin{itemize}
    \item $\M$ retriever$_{1}$ outperforms DPR by a large margin, while maintaining the same inference speed. This proves that the implicit interaction is beneficial for encoding relevance information in the final passage representation.
    \item Among the PLM-based baselines, COIL and ColBERT significantly surpass other methods. This is because
    COIL and ColBERT apply effective late interaction between the multi-vector representations of actual query and passage. However, such effectiveness costs extensive computation and storage (i.e., caching multiple vectors for each passage). Compared with COIL and ColBERT, our $\M$ retriever$_{1}$ method is more efficient, and can achieve comparable performance w.r.t. Recall@1000 on MARCO DEV Passage, and better performance w.r.t. NDCG@10 on TREC DL 19.
    \item 
    $\M$ retriever$_{1}$ is significantly better than DCE. Note that DCE also introduces interaction between pseudo-query and passage during passage encoding, where the pseudo-query is drawn from an off-the-shelf docT5query model~\cite{nogueira2019doc2query}. As such, we can conclude that the superiority of $\M$ retriever$_{1}$
    can be attributed to the joint optimization of pseudo-query reconstruction and retrieval, which makes the implicit interaction more aligned with the downstream retrieval task.
    \item 
    $\M$ retriever$_{1}$
    shows more significant improvement on TREC DL 19 than on MARCO-DEV. In particular, $\M$ retriever$_{1}$ beats all the baseline methods on TREC DL 19, including COIL and ColBERT. This implies that our proposed implicit interaction can more accurately captures fine-grained relevance ranking than the baselines.
\end{itemize}

Next, we draw more findings from the second and third groups (i.e., methods with pre-training or distillation):
\begin{itemize}
    \item $\M$ can further improve those methods that leverage special pre-training or knowledge distillation, which shows that implicit interaction is compatible with these commonly-used techniques to achieve better results.
    \item By combining implicit interaction, pre-training and distillation, $\M$ retriever$_{4}$ is able to achieve the state-of-the-art performance on both datasets and across all metrics.  
\end{itemize}

\begin{table}[t]
    \centering
    \caption{Query latency and storage cost.}
    \label{tab:efficiency}
    \begin{tabular}{l|ccc|c}
    \hline \hline \multirow{2}{*}{ Methods } & \multicolumn{3}{c|}{ \# Candidates }& \multirow{2}{*}{Space(GiBs)}\\
    \cline { 2 - 4 } & $1 \mathrm{k}$ & $100 \mathrm{k}$ &  $8.8 \mathrm{m}$ & \\
    \hline 
    Dual-encoder & \textbf{18ms} & \textbf{22ms} & \textbf{62.2ms}& \textbf{25.6} \\
    COIL & 41ms & 69ms &  344ms & 110.8\\
    ColBERT & 50ms & 83ms & 430ms & 154\\
    Cross-encoder & $2.4 \mathrm{s}$ & $4.0 \mathrm{m}$ & $5.9 \mathrm{h}$ & - \\
    \hline 
    $\M$ retriever & \textbf{18ms} & \textbf{22ms} & \textbf{62.2ms}& \textbf{25.6} \\
    \hline \hline
    \end{tabular}
\end{table}

\begin{table*}
	\caption{Performance on different groups of passages. The relative improvements are reported over dual-encoder.}
	\label{tab:frequency}
	\centering
	\begin{tabular}{l|cc|cc|cc}
		\hline \hline
		& 
        \multicolumn{2}{c|} { Overall } &
        \multicolumn{2}{c|} { Set 0 } &
        \multicolumn{2}{c} { Set 1 }\\
		Model & MRR@10 & Imp.\% & MRR@10 & Imp.\%& MRR@10 & Imp.\% \\
		\hline
        BM25         & .187 & -45.3&  .192  &-44.8& .072  &-64.2 \\
        \hline
		Dual-encoder &.342&-& .348 &-& .201 &-\\
		Cross-encoder &.399&\textbf{16.7}& .407 & \textbf{17.0} & .224 & 11.4\\
        $\M$ retriever$_2$ &.366&7.0& .372 & 6.9 & .245 & \textbf{21.9} \\
		\hline \hline
	\end{tabular}
\end{table*}

\begin{table*}
    \caption{The cases of passage with multiple relevant queries. The blue texts represent those terms that are consistent with the topic of the training query, and the red texts represent those terms that are inconsistent with the topic of the training query.}
    \label{tab:passageCase}
    \begin{tabular}{l|l|l|l}
    \hline
    \hline
        Passage & \multicolumn{3}{l}{\makecell[l]{Preheat the oven to \textcolor{blue}{450 degrees} F. Season salmon with salt and pepper. Place salmon, skin side down, on\\ a non-stick baking sheet or in a non-stick pan with an oven-proof handle. Bake until salmon is cooked t-\\hrough, about \textcolor{red}{12 to 15 minutes}.}} \\
        \hline
        \multirow{5}{*}{Relevant queries} & Training query& \multicolumn{2}{l}{best \textcolor{blue}{temperature} to cook salmon} \\ 
        \cline{2-4}
        &\multirow{4}{*}{Testing queries}&\multirow{2}{*}{best oven \textcolor{blue}{temperature} for baked salmon} & Dual-encoder ranks the passage at \#1\\
        &&&$\M$ ranks the passage at \#1 \\ 
        \cline{3-4}
        &&  \multirow{2}{*}{\textcolor{red}{how long} to cook salmon cakes in oven}
        &Dual-encoder ranks the passage at \#7\\
        &&&$\M$ ranks the passage at \#1 \\
        \hline
        Reconstructed query terms & \multicolumn{3}{l}{\textcolor{red}{how; long;} what; salmon; \textcolor{red}{minute;} \textcolor{blue}{temperature;} oven; bake; cook} \\
        \hline
        Passage &  \multicolumn{3}{l}{\makecell[l]{Tetanus, Diphtheria, Pertussis Vaccine for Adults \textcolor{blue}{tdap} is a combination \textcolor{blue}{vaccine} that protects against three \\
        potentially life-threatening bacterial diseases: tetanus, diphtheria, and pertussis (whooping cough). Td is a \\
        \textcolor{red}{booster} vaccine for tetanus and diphtheria. It does not protect against pertussis. Tetanus enters the body t-\\
        hrough a wound or cut.}}\\
        \hline

        \multirow{5}{*}{Relevant queries} & Training query& \multicolumn{2}{l}{what is a \textcolor{blue}{tdap immunization}} \\ 
        \cline{2-4}
        &\multirow{4}{*}{Testing queries}&\multirow{2}{*}{what is the \textcolor{blue}{tdap vaccine}}
        &Dual-encoder ranks the passage at \#1\\
        &&&$\M$ ranks the passage at \#1 \\  
        \cline{3-4}
        && \multirow{2}{*}{what is the tdap \textcolor{red}{booster}} & Dual-encoder ranks the passage at \#3\\
        &&&$\M$ ranks the passage at \#1 \\
        \hline
        Reconstructed query terms & \multicolumn{3}{l}{\textcolor{blue}{what; vaccine;} immunity; bacterial; immune; \textcolor{red}{booster;} diseases}\\
    \hline \hline
    \end{tabular}
\end{table*}

\noindent \textbf{Efficiency.} Tabel~\ref{tab:efficiency} shows the efficiency comparison of $\M$ and four representative models. We report the inference (i.e., relevance computation) time per query for 1,000, 100,000 and all (around 8.8 million) candidate passages as the key metrics.
First, dual-encoders are without doubt the most efficient, as no query-passage interaction is involved. All the passage representations can be pre-computed and cached, which significantly saves the inference time. Second, late-interaction encoders, such as COIL and ColBERT, usually require extra computation to perform effective late interaction during inference. 
Third, our $\M$ model is a promising solution that achieves remarkable performance on both effectiveness and efficiency. Unlike late-interaction that often undermines the inference efficiency, the implicit interaction introduced by $\M$ can be pre-computed, and the final query-aware passage representation can be cached. This allows $\M$ to be as efficient as vanilla dual-encoders. 

\subsection{Investigation on Implicit Interaction}

In this section, we investigate how the implicit interaction affects the model performance on different passages. Specifically, it worth noting that some passages (namely \textbf{Type 1} passages) have relevant queries in the training data. On the other hand, there are many other passages (namely \textbf{Type 0} passages) that do not have relevant queries in the training data.
In real-world scenarios, Type 1 passages might be those articles with abundant information that is desired by many queries, 
while Type 0 passages might be articles with specific information that can only be retrieved by a specific query.
To investigate the performance of $\M$ on the two types of passages, we divide the queries in MSMARCO DEV into two validation sets, namely \textbf{Set 0} and \textbf{Set 1}, where all the relevant passages in Set 0 are Type 0 passages, and all the relevant passages in Set 1 are Type 1 passages.
Table~\ref{tab:frequency} shows the performance comparison on MSMARCO DEV and the two divided validatation sets. We compare $\M$ retriever with our implementation of vanilla dual-encoder with the same negative sampling~\cite{xiong2020approximate}. We also include cross-encoders in the comparison, where the results are obtained by directly reranking the candidates retrieved by $\M$. We can see from the table that 
1) $\M$ can consistently outperform dual-encoders on both Set 0 and Set 1, which means that the implicit interaction is effective for both types of passages; 
2) $\M$ can achieve larger gain over dual-encoders on Set 1, and surprisingly outperform cross-encoders, which indicates that the implicit interaction is even more effective for Type 1 passages associated with multiple relevant training queries. 

\subsection{Case Study on Query Reconstruction}
To better understand the implicit interaction incorporated in $\M$, we demonstrate two cases in Table \ref{tab:passageCase}, and interpret their
query reconstruction. 
Notably, the reconstructed query terms in Table \ref{tab:passageCase} are decoded by $\mathbf{W}^{R}$ in Eq.~\ref{eq:reconstructor} and are only used for the purpose of this case study.
For each of the two passages, its query reconstruction is trained on one training query, and we presents the rankings of $\M$ and dual-encoder for two testing queries. Note that one of the testing query is similar to the training query, and the other one is dissimilar to the training query.

First, we find out that the reconstructed query terms can address several key concepts and terms that a query might ask for in a long passage. This indicates that our implicit interaction can help identifying important concepts during passage encoding and eventually boosting the final performance. This finding also justifies the results presented in Table \ref{tab:frequency}, where the relative improvement of $\M$ over dual-encoder is larger on Type 1 passages. Second, it worth noting that the reconstructed query terms are not just the memorization of training query. In fact, they are also generalized to the key terms that are not covered by the training query. For example, the first passage contains information about the temperature and the time of cooking salmon. We note that both two aspects are able to be covered by the reconstructed query terms, while the model is only trained on the training query that asks for temperature. As such, both dual-encoder and $\M$ can perform well on the testing query that is similar to the training query (i.e., ranking the passage at \#1), while $\M$ performs much better on the testing query that is dissimilar to the training query. This concludes that the generalization ability of extracting key concepts of passages might be the key of the success of $\M$.

\section{Conclusion}
In this paper, we propose a new interaction paradigm for dense retrieval, namely $\M$ retriever, which incorporates implicit interaction into dual-encoders. 
Particularly, $\M$ advances conventional dual-encoders with
1) a lightweight query reconstructor 
and 2) a query-passage interactor, which generate pseudo-query for expressive interaction.
By doing this, our $\M$ model is equipped with the capability of modeling implicit interaction, leading to an effective and efficient encoding of semantic relevance features in the final passage representations.
The evaluation shows that the retrieval performance could be significantly improved without introducing extra computational overhead and space footprint. 
Besides, we also show that the proposed implicit interaction is compatible with special pretraining and distillation to achieve a better performance.

\begin{acks}
This work is supported by Quan Cheng Laboratory (Grant No. QCLZD202301).
\end{acks}
\bibliographystyle{ACM-Reference-Format}
\balance
\bibliography{sample-base}


\begin{thebibliography}{68}


\ifx \showCODEN    \undefined \def \showCODEN     #1{\unskip}     \fi
\ifx \showDOI      \undefined \def \showDOI       #1{#1}\fi
\ifx \showISBNx    \undefined \def \showISBNx     #1{\unskip}     \fi
\ifx \showISBNxiii \undefined \def \showISBNxiii  #1{\unskip}     \fi
\ifx \showISSN     \undefined \def \showISSN      #1{\unskip}     \fi
\ifx \showLCCN     \undefined \def \showLCCN      #1{\unskip}     \fi
\ifx \shownote     \undefined \def \shownote      #1{#1}          \fi
\ifx \showarticletitle \undefined \def \showarticletitle #1{#1}   \fi
\ifx \showURL      \undefined \def \showURL       {\relax}        \fi
\providecommand\bibfield[2]{#2}
\providecommand\bibinfo[2]{#2}
\providecommand\natexlab[1]{#1}
\providecommand\showeprint[2][]{arXiv:#2}

\bibitem[Bonifacio et~al\mbox{.}(2022)]%
        {bonifacio2022inpars}
\bibfield{author}{\bibinfo{person}{Luiz Bonifacio}, \bibinfo{person}{Hugo
  Abonizio}, \bibinfo{person}{Marzieh Fadaee}, {and} \bibinfo{person}{Rodrigo
  Nogueira}.} \bibinfo{year}{2022}\natexlab{}.
\newblock \showarticletitle{InPars: Data Augmentation for Information Retrieval
  using Large Language Models}.
\newblock \bibinfo{journal}{\emph{arXiv preprint arXiv:2202.05144}}
  (\bibinfo{year}{2022}).
\newblock


\bibitem[Chang et~al\mbox{.}(2020)]%
        {chang2020pre}
\bibfield{author}{\bibinfo{person}{Wei-Cheng Chang}, \bibinfo{person}{Felix~X
  Yu}, \bibinfo{person}{Yin-Wen Chang}, \bibinfo{person}{Yiming Yang}, {and}
  \bibinfo{person}{Sanjiv Kumar}.} \bibinfo{year}{2020}\natexlab{}.
\newblock \showarticletitle{Pre-training tasks for embedding-based large-scale
  retrieval}.
\newblock \bibinfo{journal}{\emph{arXiv preprint arXiv:2002.03932}}
  (\bibinfo{year}{2020}).
\newblock


\bibitem[Cheng et~al\mbox{.}(2023)]%
        {cheng2023layout}
\bibfield{author}{\bibinfo{person}{Anfeng Cheng}, \bibinfo{person}{Yiding Liu},
  \bibinfo{person}{Weibin Li}, \bibinfo{person}{Qian Dong},
  \bibinfo{person}{Shuaiqiang Wang}, \bibinfo{person}{Zhengjie Huang},
  \bibinfo{person}{Shikun Feng}, \bibinfo{person}{Zhicong Cheng}, {and}
  \bibinfo{person}{Dawei Yin}.} \bibinfo{year}{2023}\natexlab{}.
\newblock \showarticletitle{Layout-aware Webpage Quality Assessment}.
\newblock \bibinfo{journal}{\emph{arXiv preprint arXiv:2301.12152}}
  (\bibinfo{year}{2023}).
\newblock


\bibitem[Chung et~al\mbox{.}(2022)]%
        {chung2022scaling}
\bibfield{author}{\bibinfo{person}{Hyung~Won Chung}, \bibinfo{person}{Le Hou},
  \bibinfo{person}{Shayne Longpre}, \bibinfo{person}{Barret Zoph},
  \bibinfo{person}{Yi Tay}, \bibinfo{person}{William Fedus},
  \bibinfo{person}{Eric Li}, \bibinfo{person}{Xuezhi Wang},
  \bibinfo{person}{Mostafa Dehghani}, \bibinfo{person}{Siddhartha Brahma},
  {et~al\mbox{.}}} \bibinfo{year}{2022}\natexlab{}.
\newblock \showarticletitle{Scaling instruction-finetuned language models}.
\newblock \bibinfo{journal}{\emph{arXiv preprint arXiv:2210.11416}}
  (\bibinfo{year}{2022}).
\newblock


\bibitem[Craswell et~al\mbox{.}(2020)]%
        {craswell2020overview}
\bibfield{author}{\bibinfo{person}{Nick Craswell}, \bibinfo{person}{Bhaskar
  Mitra}, \bibinfo{person}{Emine Yilmaz}, \bibinfo{person}{Daniel Campos},
  {and} \bibinfo{person}{Ellen~M Voorhees}.} \bibinfo{year}{2020}\natexlab{}.
\newblock \showarticletitle{Overview of the trec 2019 deep learning track}.
\newblock \bibinfo{journal}{\emph{arXiv preprint arXiv:2003.07820}}
  (\bibinfo{year}{2020}).
\newblock


\bibitem[Dai and Callan(2020)]%
        {dai2020context}
\bibfield{author}{\bibinfo{person}{Zhuyun Dai} {and} \bibinfo{person}{Jamie
  Callan}.} \bibinfo{year}{2020}\natexlab{}.
\newblock \showarticletitle{Context-aware term weighting for first stage
  passage retrieval}. In \bibinfo{booktitle}{\emph{Proceedings of the 43rd
  International ACM SIGIR conference on research and development in Information
  Retrieval}}. \bibinfo{pages}{1533--1536}.
\newblock


\bibitem[Dai et~al\mbox{.}(2018)]%
        {dai2018convolutional}
\bibfield{author}{\bibinfo{person}{Zhuyun Dai}, \bibinfo{person}{Chenyan
  Xiong}, \bibinfo{person}{Jamie Callan}, {and} \bibinfo{person}{Zhiyuan Liu}.}
  \bibinfo{year}{2018}\natexlab{}.
\newblock \showarticletitle{Convolutional neural networks for soft-matching
  n-grams in ad-hoc search}. In \bibinfo{booktitle}{\emph{Proceedings of the
  eleventh ACM international conference on web search and data mining}}.
  \bibinfo{pages}{126--134}.
\newblock


\bibitem[Dai et~al\mbox{.}(2022)]%
        {dai2022promptagator}
\bibfield{author}{\bibinfo{person}{Zhuyun Dai}, \bibinfo{person}{Vincent~Y
  Zhao}, \bibinfo{person}{Ji Ma}, \bibinfo{person}{Yi Luan},
  \bibinfo{person}{Jianmo Ni}, \bibinfo{person}{Jing Lu},
  \bibinfo{person}{Anton Bakalov}, \bibinfo{person}{Kelvin Guu},
  \bibinfo{person}{Keith~B Hall}, {and} \bibinfo{person}{Ming-Wei Chang}.}
  \bibinfo{year}{2022}\natexlab{}.
\newblock \showarticletitle{Promptagator: Few-shot dense retrieval from 8
  examples}.
\newblock \bibinfo{journal}{\emph{arXiv preprint arXiv:2209.11755}}
  (\bibinfo{year}{2022}).
\newblock


\bibitem[Devlin et~al\mbox{.}(2018)]%
        {devlin2018bert}
\bibfield{author}{\bibinfo{person}{Jacob Devlin}, \bibinfo{person}{Ming-Wei
  Chang}, \bibinfo{person}{Kenton Lee}, {and} \bibinfo{person}{Kristina
  Toutanova}.} \bibinfo{year}{2018}\natexlab{}.
\newblock \showarticletitle{Bert: Pre-training of deep bidirectional
  transformers for language understanding}.
\newblock \bibinfo{journal}{\emph{arXiv preprint arXiv:1810.04805}}
  (\bibinfo{year}{2018}).
\newblock


\bibitem[Dong et~al\mbox{.}(2022a)]%
        {dong2022incorporating}
\bibfield{author}{\bibinfo{person}{Qian Dong}, \bibinfo{person}{Yiding Liu},
  \bibinfo{person}{Suqi Cheng}, \bibinfo{person}{Shuaiqiang Wang},
  \bibinfo{person}{Zhicong Cheng}, \bibinfo{person}{Shuzi Niu}, {and}
  \bibinfo{person}{Dawei Yin}.} \bibinfo{year}{2022}\natexlab{a}.
\newblock \showarticletitle{Incorporating Explicit Knowledge in Pre-trained
  Language Models for Passage Re-ranking}.
\newblock \bibinfo{journal}{\emph{arXiv preprint arXiv:2204.11673}}
  (\bibinfo{year}{2022}).
\newblock


\bibitem[Dong and Niu(2021a)]%
        {dong2021latent}
\bibfield{author}{\bibinfo{person}{Qian Dong} {and} \bibinfo{person}{Shuzi
  Niu}.} \bibinfo{year}{2021}\natexlab{a}.
\newblock \showarticletitle{Latent Graph Recurrent Network for Document
  Ranking}. In \bibinfo{booktitle}{\emph{Database Systems for Advanced
  Applications: 26th International Conference, DASFAA 2021, Taipei, Taiwan,
  April 11--14, 2021, Proceedings, Part II 26}}. Springer,
  \bibinfo{pages}{88--103}.
\newblock


\bibitem[Dong and Niu(2021b)]%
        {dong2021legal}
\bibfield{author}{\bibinfo{person}{Qian Dong} {and} \bibinfo{person}{Shuzi
  Niu}.} \bibinfo{year}{2021}\natexlab{b}.
\newblock \showarticletitle{Legal judgment prediction via relational learning}.
  In \bibinfo{booktitle}{\emph{Proceedings of the 44th International ACM SIGIR
  Conference on Research and Development in Information Retrieval}}.
  \bibinfo{pages}{983--992}.
\newblock


\bibitem[Dong et~al\mbox{.}(2022b)]%
        {dong2022disentangled}
\bibfield{author}{\bibinfo{person}{Qian Dong}, \bibinfo{person}{Shuzi Niu},
  \bibinfo{person}{Tao Yuan}, {and} \bibinfo{person}{Yucheng Li}.}
  \bibinfo{year}{2022}\natexlab{b}.
\newblock \showarticletitle{Disentangled graph recurrent network for document
  ranking}.
\newblock \bibinfo{journal}{\emph{Data Science and Engineering}}
  \bibinfo{volume}{7}, \bibinfo{number}{1} (\bibinfo{year}{2022}),
  \bibinfo{pages}{30--43}.
\newblock


\bibitem[Formal et~al\mbox{.}(2021)]%
        {formal2021splade}
\bibfield{author}{\bibinfo{person}{Thibault Formal}, \bibinfo{person}{Carlos
  Lassance}, \bibinfo{person}{Benjamin Piwowarski}, {and}
  \bibinfo{person}{St{\'e}phane Clinchant}.} \bibinfo{year}{2021}\natexlab{}.
\newblock \showarticletitle{SPLADE v2: Sparse lexical and expansion model for
  information retrieval}.
\newblock \bibinfo{journal}{\emph{arXiv preprint arXiv:2109.10086}}
  (\bibinfo{year}{2021}).
\newblock


\bibitem[Gao and Callan(2021)]%
        {gao2021unsupervised}
\bibfield{author}{\bibinfo{person}{Luyu Gao} {and} \bibinfo{person}{Jamie
  Callan}.} \bibinfo{year}{2021}\natexlab{}.
\newblock \showarticletitle{Unsupervised corpus aware language model
  pre-training for dense passage retrieval}.
\newblock \bibinfo{journal}{\emph{arXiv preprint arXiv:2108.05540}}
  (\bibinfo{year}{2021}).
\newblock


\bibitem[Gao et~al\mbox{.}(2021)]%
        {gao2021coil}
\bibfield{author}{\bibinfo{person}{Luyu Gao}, \bibinfo{person}{Zhuyun Dai},
  {and} \bibinfo{person}{Jamie Callan}.} \bibinfo{year}{2021}\natexlab{}.
\newblock \showarticletitle{COIL: Revisit exact lexical match in information
  retrieval with contextualized inverted list}.
\newblock \bibinfo{journal}{\emph{arXiv preprint arXiv:2104.07186}}
  (\bibinfo{year}{2021}).
\newblock


\bibitem[Guo et~al\mbox{.}(2016)]%
        {guo2016deep}
\bibfield{author}{\bibinfo{person}{Jiafeng Guo}, \bibinfo{person}{Yixing Fan},
  \bibinfo{person}{Qingyao Ai}, {and} \bibinfo{person}{W~Bruce Croft}.}
  \bibinfo{year}{2016}\natexlab{}.
\newblock \showarticletitle{A deep relevance matching model for ad-hoc
  retrieval}. In \bibinfo{booktitle}{\emph{Proceedings of the 25th ACM
  international on conference on information and knowledge management}}.
  \bibinfo{pages}{55--64}.
\newblock


\bibitem[Guo et~al\mbox{.}(2020)]%
        {guo2020deep}
\bibfield{author}{\bibinfo{person}{Jiafeng Guo}, \bibinfo{person}{Yixing Fan},
  \bibinfo{person}{Liang Pang}, \bibinfo{person}{Liu Yang},
  \bibinfo{person}{Qingyao Ai}, \bibinfo{person}{Hamed Zamani},
  \bibinfo{person}{Chen Wu}, \bibinfo{person}{W~Bruce Croft}, {and}
  \bibinfo{person}{Xueqi Cheng}.} \bibinfo{year}{2020}\natexlab{}.
\newblock \showarticletitle{A deep look into neural ranking models for
  information retrieval}.
\newblock \bibinfo{journal}{\emph{Information Processing \& Management}}
  \bibinfo{volume}{57}, \bibinfo{number}{6} (\bibinfo{year}{2020}),
  \bibinfo{pages}{102067}.
\newblock


\bibitem[Hofst{\"a}tter et~al\mbox{.}(2021)]%
        {hofstatter2021efficiently}
\bibfield{author}{\bibinfo{person}{Sebastian Hofst{\"a}tter},
  \bibinfo{person}{Sheng-Chieh Lin}, \bibinfo{person}{Jheng-Hong Yang},
  \bibinfo{person}{Jimmy Lin}, {and} \bibinfo{person}{Allan Hanbury}.}
  \bibinfo{year}{2021}\natexlab{}.
\newblock \showarticletitle{Efficiently teaching an effective dense retriever
  with balanced topic aware sampling}. In \bibinfo{booktitle}{\emph{Proceedings
  of the 44th International ACM SIGIR Conference on Research and Development in
  Information Retrieval}}. \bibinfo{pages}{113--122}.
\newblock


\bibitem[Hu et~al\mbox{.}(2014)]%
        {hu2014convolutional}
\bibfield{author}{\bibinfo{person}{Baotian Hu}, \bibinfo{person}{Zhengdong Lu},
  \bibinfo{person}{Hang Li}, {and} \bibinfo{person}{Qingcai Chen}.}
  \bibinfo{year}{2014}\natexlab{}.
\newblock \showarticletitle{Convolutional neural network architectures for
  matching natural language sentences}.
\newblock \bibinfo{journal}{\emph{Advances in neural information processing
  systems}}  \bibinfo{volume}{27} (\bibinfo{year}{2014}).
\newblock


\bibitem[Huang et~al\mbox{.}(2013)]%
        {huang2013learning}
\bibfield{author}{\bibinfo{person}{Po-Sen Huang}, \bibinfo{person}{Xiaodong
  He}, \bibinfo{person}{Jianfeng Gao}, \bibinfo{person}{Li Deng},
  \bibinfo{person}{Alex Acero}, {and} \bibinfo{person}{Larry Heck}.}
  \bibinfo{year}{2013}\natexlab{}.
\newblock \showarticletitle{Learning deep structured semantic models for web
  search using clickthrough data}. In \bibinfo{booktitle}{\emph{Proceedings of
  the 22nd ACM international conference on Information \& Knowledge
  Management}}. \bibinfo{pages}{2333--2338}.
\newblock


\bibitem[Humeau et~al\mbox{.}(2019)]%
        {humeau2019poly}
\bibfield{author}{\bibinfo{person}{Samuel Humeau}, \bibinfo{person}{Kurt
  Shuster}, \bibinfo{person}{Marie-Anne Lachaux}, {and} \bibinfo{person}{Jason
  Weston}.} \bibinfo{year}{2019}\natexlab{}.
\newblock \showarticletitle{Poly-encoders: Transformer architectures and
  pre-training strategies for fast and accurate multi-sentence scoring}.
\newblock \bibinfo{journal}{\emph{arXiv preprint arXiv:1905.01969}}
  (\bibinfo{year}{2019}).
\newblock


\bibitem[Johnson et~al\mbox{.}(2019)]%
        {johnson2019billion}
\bibfield{author}{\bibinfo{person}{Jeff Johnson}, \bibinfo{person}{Matthijs
  Douze}, {and} \bibinfo{person}{Herv{\'e} J{\'e}gou}.}
  \bibinfo{year}{2019}\natexlab{}.
\newblock \showarticletitle{Billion-scale similarity search with {GPUs}}.
\newblock \bibinfo{journal}{\emph{IEEE Transactions on Big Data}}
  \bibinfo{volume}{7}, \bibinfo{number}{3} (\bibinfo{year}{2019}),
  \bibinfo{pages}{535--547}.
\newblock


\bibitem[Karpukhin et~al\mbox{.}(2020)]%
        {karpukhin2020dense}
\bibfield{author}{\bibinfo{person}{Vladimir Karpukhin}, \bibinfo{person}{Barlas
  O{\u{g}}uz}, \bibinfo{person}{Sewon Min}, \bibinfo{person}{Patrick Lewis},
  \bibinfo{person}{Ledell Wu}, \bibinfo{person}{Sergey Edunov},
  \bibinfo{person}{Danqi Chen}, {and} \bibinfo{person}{Wen-tau Yih}.}
  \bibinfo{year}{2020}\natexlab{}.
\newblock \showarticletitle{Dense passage retrieval for open-domain question
  answering}.
\newblock \bibinfo{journal}{\emph{arXiv preprint arXiv:2004.04906}}
  (\bibinfo{year}{2020}).
\newblock


\bibitem[Khattab and Zaharia(2020)]%
        {khattab2020colbert}
\bibfield{author}{\bibinfo{person}{Omar Khattab} {and} \bibinfo{person}{Matei
  Zaharia}.} \bibinfo{year}{2020}\natexlab{}.
\newblock \showarticletitle{Colbert: Efficient and effective passage search via
  contextualized late interaction over bert}. In
  \bibinfo{booktitle}{\emph{Proceedings of the 43rd International ACM SIGIR
  conference on research and development in Information Retrieval}}.
  \bibinfo{pages}{39--48}.
\newblock


\bibitem[Li et~al\mbox{.}(2023c)]%
        {li2023pretrained}
\bibfield{author}{\bibinfo{person}{Canjia Li}, \bibinfo{person}{Xiaoyang Wang},
  \bibinfo{person}{Dongdong Li}, \bibinfo{person}{Yiding Liu},
  \bibinfo{person}{Yu Lu}, \bibinfo{person}{Shuaiqiang Wang},
  \bibinfo{person}{Zhicong Cheng}, \bibinfo{person}{Simiu Gu}, {and}
  \bibinfo{person}{Dawei Yin}.} \bibinfo{year}{2023}\natexlab{c}.
\newblock \showarticletitle{Pretrained Language Model based Web Search Ranking:
  From Relevance to Satisfaction}.
\newblock \bibinfo{journal}{\emph{arXiv preprint arXiv:2306.01599}}
  (\bibinfo{year}{2023}).
\newblock


\bibitem[Li et~al\mbox{.}(2023a)]%
        {li2023sailer}
\bibfield{author}{\bibinfo{person}{Haitao Li}, \bibinfo{person}{Qingyao Ai},
  \bibinfo{person}{Jia Chen}, \bibinfo{person}{Qian Dong},
  \bibinfo{person}{Yueyue Wu}, \bibinfo{person}{Yiqun Liu},
  \bibinfo{person}{Chong Chen}, {and} \bibinfo{person}{Qi Tian}.}
  \bibinfo{year}{2023}\natexlab{a}.
\newblock \showarticletitle{SAILER: Structure-aware Pre-trained Language Model
  for Legal Case Retrieval}.
\newblock \bibinfo{journal}{\emph{arXiv preprint arXiv:2304.11370}}
  (\bibinfo{year}{2023}).
\newblock


\bibitem[Li et~al\mbox{.}(2023b)]%
        {li2023constructing}
\bibfield{author}{\bibinfo{person}{Haitao Li}, \bibinfo{person}{Qingyao Ai},
  \bibinfo{person}{Jingtao Zhan}, \bibinfo{person}{Jiaxin Mao},
  \bibinfo{person}{Yiqun Liu}, \bibinfo{person}{Zheng Liu}, {and}
  \bibinfo{person}{Zhao Cao}.} \bibinfo{year}{2023}\natexlab{b}.
\newblock \showarticletitle{Constructing Tree-based Index for Efficient and
  Effective Dense Retrieval}.
\newblock \bibinfo{journal}{\emph{arXiv preprint arXiv:2304.11943}}
  (\bibinfo{year}{2023}).
\newblock


\bibitem[Li et~al\mbox{.}(2022)]%
        {li2022learning}
\bibfield{author}{\bibinfo{person}{Zehan Li}, \bibinfo{person}{Nan Yang},
  \bibinfo{person}{Liang Wang}, {and} \bibinfo{person}{Furu Wei}.}
  \bibinfo{year}{2022}\natexlab{}.
\newblock \showarticletitle{Learning Diverse Document Representations with Deep
  Query Interactions for Dense Retrieval}.
\newblock \bibinfo{journal}{\emph{arXiv preprint arXiv:2208.04232}}
  (\bibinfo{year}{2022}).
\newblock


\bibitem[Liang et~al\mbox{.}(2020)]%
        {liang2020embedding}
\bibfield{author}{\bibinfo{person}{Davis Liang}, \bibinfo{person}{Peng Xu},
  \bibinfo{person}{Siamak Shakeri}, \bibinfo{person}{Cicero Nogueira~dos
  Santos}, \bibinfo{person}{Ramesh Nallapati}, \bibinfo{person}{Zhiheng Huang},
  {and} \bibinfo{person}{Bing Xiang}.} \bibinfo{year}{2020}\natexlab{}.
\newblock \showarticletitle{Embedding-based zero-shot retrieval through query
  generation}.
\newblock \bibinfo{journal}{\emph{arXiv preprint arXiv:2009.10270}}
  (\bibinfo{year}{2020}).
\newblock


\bibitem[Liu et~al\mbox{.}(2019)]%
        {liu2019roberta}
\bibfield{author}{\bibinfo{person}{Yinhan Liu}, \bibinfo{person}{Myle Ott},
  \bibinfo{person}{Naman Goyal}, \bibinfo{person}{Jingfei Du},
  \bibinfo{person}{Mandar Joshi}, \bibinfo{person}{Danqi Chen},
  \bibinfo{person}{Omer Levy}, \bibinfo{person}{Mike Lewis},
  \bibinfo{person}{Luke Zettlemoyer}, {and} \bibinfo{person}{Veselin
  Stoyanov}.} \bibinfo{year}{2019}\natexlab{}.
\newblock \showarticletitle{Roberta: A robustly optimized bert pretraining
  approach}.
\newblock \bibinfo{journal}{\emph{arXiv preprint arXiv:1907.11692}}
  (\bibinfo{year}{2019}).
\newblock


\bibitem[Liu and Shao(2022)]%
        {liu2022retromae}
\bibfield{author}{\bibinfo{person}{Zheng Liu} {and} \bibinfo{person}{Yingxia
  Shao}.} \bibinfo{year}{2022}\natexlab{}.
\newblock \showarticletitle{Retromae: Pre-training retrieval-oriented
  transformers via masked auto-encoder}.
\newblock \bibinfo{journal}{\emph{arXiv preprint arXiv:2205.12035}}
  (\bibinfo{year}{2022}).
\newblock


\bibitem[Lu et~al\mbox{.}(2021)]%
        {lu2021less}
\bibfield{author}{\bibinfo{person}{Shuqi Lu}, \bibinfo{person}{Di He},
  \bibinfo{person}{Chenyan Xiong}, \bibinfo{person}{Guolin Ke},
  \bibinfo{person}{Waleed Malik}, \bibinfo{person}{Zhicheng Dou},
  \bibinfo{person}{Paul Bennett}, \bibinfo{person}{Tie-Yan Liu}, {and}
  \bibinfo{person}{Arnold Overwijk}.} \bibinfo{year}{2021}\natexlab{}.
\newblock \showarticletitle{Less is More: Pretrain a Strong Siamese Encoder for
  Dense Text Retrieval Using a Weak Decoder}. In
  \bibinfo{booktitle}{\emph{Proceedings of the 2021 Conference on Empirical
  Methods in Natural Language Processing}}. \bibinfo{pages}{2780--2791}.
\newblock


\bibitem[Lu et~al\mbox{.}(2022)]%
        {lu2022ernie}
\bibfield{author}{\bibinfo{person}{Yuxiang Lu}, \bibinfo{person}{Yiding Liu},
  \bibinfo{person}{Jiaxiang Liu}, \bibinfo{person}{Yunsheng Shi},
  \bibinfo{person}{Zhengjie Huang}, \bibinfo{person}{Shikun Feng~Yu Sun},
  \bibinfo{person}{Hao Tian}, \bibinfo{person}{Hua Wu},
  \bibinfo{person}{Shuaiqiang Wang}, \bibinfo{person}{Dawei Yin},
  {et~al\mbox{.}}} \bibinfo{year}{2022}\natexlab{}.
\newblock \showarticletitle{Ernie-search: Bridging cross-encoder with
  dual-encoder via self on-the-fly distillation for dense passage retrieval}.
\newblock \bibinfo{journal}{\emph{arXiv preprint arXiv:2205.09153}}
  (\bibinfo{year}{2022}).
\newblock


\bibitem[Luan et~al\mbox{.}(2021)]%
        {luan2021sparse}
\bibfield{author}{\bibinfo{person}{Yi Luan}, \bibinfo{person}{Jacob
  Eisenstein}, \bibinfo{person}{Kristina Toutanova}, {and}
  \bibinfo{person}{Michael Collins}.} \bibinfo{year}{2021}\natexlab{}.
\newblock \showarticletitle{Sparse, dense, and attentional representations for
  text retrieval}.
\newblock \bibinfo{journal}{\emph{Transactions of the Association for
  Computational Linguistics}}  \bibinfo{volume}{9} (\bibinfo{year}{2021}),
  \bibinfo{pages}{329--345}.
\newblock


\bibitem[Ma et~al\mbox{.}(2020)]%
        {ma2020zero}
\bibfield{author}{\bibinfo{person}{Ji Ma}, \bibinfo{person}{Ivan Korotkov},
  \bibinfo{person}{Yinfei Yang}, \bibinfo{person}{Keith Hall}, {and}
  \bibinfo{person}{Ryan McDonald}.} \bibinfo{year}{2020}\natexlab{}.
\newblock \showarticletitle{Zero-shot neural passage retrieval via
  domain-targeted synthetic question generation}.
\newblock \bibinfo{journal}{\emph{arXiv preprint arXiv:2004.14503}}
  (\bibinfo{year}{2020}).
\newblock


\bibitem[Ma et~al\mbox{.}(2022)]%
        {ma2022pre}
\bibfield{author}{\bibinfo{person}{Xinyu Ma}, \bibinfo{person}{Jiafeng Guo},
  \bibinfo{person}{Ruqing Zhang}, \bibinfo{person}{Yixing Fan}, {and}
  \bibinfo{person}{Xueqi Cheng}.} \bibinfo{year}{2022}\natexlab{}.
\newblock \showarticletitle{Pre-train a Discriminative Text Encoder for Dense
  Retrieval via Contrastive Span Prediction}.
\newblock \bibinfo{journal}{\emph{arXiv preprint arXiv:2204.10641}}
  (\bibinfo{year}{2022}).
\newblock


\bibitem[Malkov and Yashunin(2018)]%
        {malkov2018efficient}
\bibfield{author}{\bibinfo{person}{Yu~A Malkov} {and} \bibinfo{person}{Dmitry~A
  Yashunin}.} \bibinfo{year}{2018}\natexlab{}.
\newblock \showarticletitle{Efficient and robust approximate nearest neighbor
  search using hierarchical navigable small world graphs}.
\newblock \bibinfo{journal}{\emph{IEEE transactions on pattern analysis and
  machine intelligence}} \bibinfo{volume}{42}, \bibinfo{number}{4}
  (\bibinfo{year}{2018}), \bibinfo{pages}{824--836}.
\newblock


\bibitem[Nguyen et~al\mbox{.}(2016)]%
        {nguyen2016ms}
\bibfield{author}{\bibinfo{person}{Tri Nguyen}, \bibinfo{person}{Mir
  Rosenberg}, \bibinfo{person}{Xia Song}, \bibinfo{person}{Jianfeng Gao},
  \bibinfo{person}{Saurabh Tiwary}, \bibinfo{person}{Rangan Majumder}, {and}
  \bibinfo{person}{Li Deng}.} \bibinfo{year}{2016}\natexlab{}.
\newblock \showarticletitle{MS MARCO: A human generated machine reading
  comprehension dataset}. In \bibinfo{booktitle}{\emph{CoCo@ NIPS}}.
\newblock


\bibitem[Nogueira and Cho(2019)]%
        {nogueira2019passage}
\bibfield{author}{\bibinfo{person}{Rodrigo Nogueira} {and}
  \bibinfo{person}{Kyunghyun Cho}.} \bibinfo{year}{2019}\natexlab{}.
\newblock \showarticletitle{Passage Re-ranking with BERT}.
\newblock \bibinfo{journal}{\emph{arXiv preprint arXiv:1901.04085}}
  (\bibinfo{year}{2019}).
\newblock


\bibitem[Nogueira et~al\mbox{.}(2019a)]%
        {nogueira2019doc2query}
\bibfield{author}{\bibinfo{person}{Rodrigo Nogueira}, \bibinfo{person}{Jimmy
  Lin}, {and} \bibinfo{person}{AI Epistemic}.}
  \bibinfo{year}{2019}\natexlab{a}.
\newblock \showarticletitle{From doc2query to docTTTTTquery}.
\newblock \bibinfo{journal}{\emph{Online preprint}}  \bibinfo{volume}{6}
  (\bibinfo{year}{2019}).
\newblock


\bibitem[Nogueira et~al\mbox{.}(2019b)]%
        {nogueira2019multi}
\bibfield{author}{\bibinfo{person}{Rodrigo Nogueira}, \bibinfo{person}{Wei
  Yang}, \bibinfo{person}{Kyunghyun Cho}, {and} \bibinfo{person}{Jimmy Lin}.}
  \bibinfo{year}{2019}\natexlab{b}.
\newblock \showarticletitle{Multi-stage document ranking with bert}.
\newblock \bibinfo{journal}{\emph{arXiv preprint arXiv:1910.14424}}
  (\bibinfo{year}{2019}).
\newblock


\bibitem[Nogueira et~al\mbox{.}(2019c)]%
        {nogueira2019document}
\bibfield{author}{\bibinfo{person}{Rodrigo Nogueira}, \bibinfo{person}{Wei
  Yang}, \bibinfo{person}{Jimmy Lin}, {and} \bibinfo{person}{Kyunghyun Cho}.}
  \bibinfo{year}{2019}\natexlab{c}.
\newblock \showarticletitle{Document expansion by query prediction}.
\newblock \bibinfo{journal}{\emph{arXiv preprint arXiv:1904.08375}}
  (\bibinfo{year}{2019}).
\newblock


\bibitem[Palangi et~al\mbox{.}(2016)]%
        {palangi2016deep}
\bibfield{author}{\bibinfo{person}{Hamid Palangi}, \bibinfo{person}{Li Deng},
  \bibinfo{person}{Yelong Shen}, \bibinfo{person}{Jianfeng Gao},
  \bibinfo{person}{Xiaodong He}, \bibinfo{person}{Jianshu Chen},
  \bibinfo{person}{Xinying Song}, {and} \bibinfo{person}{Rabab Ward}.}
  \bibinfo{year}{2016}\natexlab{}.
\newblock \showarticletitle{Deep sentence embedding using long short-term
  memory networks: Analysis and application to information retrieval}.
\newblock \bibinfo{journal}{\emph{IEEE/ACM Transactions on Audio, Speech, and
  Language Processing}} \bibinfo{volume}{24}, \bibinfo{number}{4}
  (\bibinfo{year}{2016}), \bibinfo{pages}{694--707}.
\newblock


\bibitem[Qiu and Huang(2015)]%
        {qiu2015convolutional}
\bibfield{author}{\bibinfo{person}{Xipeng Qiu} {and} \bibinfo{person}{Xuanjing
  Huang}.} \bibinfo{year}{2015}\natexlab{}.
\newblock \showarticletitle{Convolutional neural tensor network architecture
  for community-based question answering}. In
  \bibinfo{booktitle}{\emph{Twenty-Fourth international joint conference on
  artificial intelligence}}.
\newblock


\bibitem[Qu et~al\mbox{.}(2021)]%
        {qu2021rocketqa}
\bibfield{author}{\bibinfo{person}{Yingqi Qu}, \bibinfo{person}{Yuchen Ding},
  \bibinfo{person}{Jing Liu}, \bibinfo{person}{Kai Liu},
  \bibinfo{person}{Ruiyang Ren}, \bibinfo{person}{Wayne~Xin Zhao},
  \bibinfo{person}{Daxiang Dong}, \bibinfo{person}{Hua Wu}, {and}
  \bibinfo{person}{Haifeng Wang}.} \bibinfo{year}{2021}\natexlab{}.
\newblock \showarticletitle{RocketQA: An optimized training approach to dense
  passage retrieval for open-domain question answering}. In
  \bibinfo{booktitle}{\emph{Proceedings of the 2021 Conference of the North
  American Chapter of the Association for Computational Linguistics: Human
  Language Technologies}}. \bibinfo{pages}{5835--5847}.
\newblock


\bibitem[Radford et~al\mbox{.}(2018)]%
        {radford2018improving}
\bibfield{author}{\bibinfo{person}{Alec Radford}, \bibinfo{person}{Karthik
  Narasimhan}, \bibinfo{person}{Tim Salimans}, \bibinfo{person}{Ilya
  Sutskever}, {et~al\mbox{.}}} \bibinfo{year}{2018}\natexlab{}.
\newblock \showarticletitle{Improving language understanding by generative
  pre-training}.
\newblock  (\bibinfo{year}{2018}).
\newblock


\bibitem[Raffel et~al\mbox{.}(2020)]%
        {raffel2020exploring}
\bibfield{author}{\bibinfo{person}{Colin Raffel}, \bibinfo{person}{Noam
  Shazeer}, \bibinfo{person}{Adam Roberts}, \bibinfo{person}{Katherine Lee},
  \bibinfo{person}{Sharan Narang}, \bibinfo{person}{Michael Matena},
  \bibinfo{person}{Yanqi Zhou}, \bibinfo{person}{Wei Li},
  \bibinfo{person}{Peter~J Liu}, {et~al\mbox{.}}}
  \bibinfo{year}{2020}\natexlab{}.
\newblock \showarticletitle{Exploring the limits of transfer learning with a
  unified text-to-text transformer.}
\newblock \bibinfo{journal}{\emph{J. Mach. Learn. Res.}} \bibinfo{volume}{21},
  \bibinfo{number}{140} (\bibinfo{year}{2020}), \bibinfo{pages}{1--67}.
\newblock


\bibitem[Ren et~al\mbox{.}(2021)]%
        {ren2021rocketqav2}
\bibfield{author}{\bibinfo{person}{Ruiyang Ren}, \bibinfo{person}{Yingqi Qu},
  \bibinfo{person}{Jing Liu}, \bibinfo{person}{Wayne~Xin Zhao},
  \bibinfo{person}{Qiaoqiao She}, \bibinfo{person}{Hua Wu},
  \bibinfo{person}{Haifeng Wang}, {and} \bibinfo{person}{Ji-Rong Wen}.}
  \bibinfo{year}{2021}\natexlab{}.
\newblock \showarticletitle{RocketQAv2: A Joint Training Method for Dense
  Passage Retrieval and Passage Re-ranking}.
\newblock \bibinfo{journal}{\emph{arXiv preprint arXiv:2110.07367}}
  (\bibinfo{year}{2021}).
\newblock


\bibitem[Robertson et~al\mbox{.}(2009)]%
        {robertson2009probabilistic}
\bibfield{author}{\bibinfo{person}{Stephen Robertson}, \bibinfo{person}{Hugo
  Zaragoza}, {et~al\mbox{.}}} \bibinfo{year}{2009}\natexlab{}.
\newblock \showarticletitle{The probabilistic relevance framework: BM25 and
  beyond}.
\newblock \bibinfo{journal}{\emph{Foundations and Trends{\textregistered} in
  Information Retrieval}} \bibinfo{volume}{3}, \bibinfo{number}{4}
  (\bibinfo{year}{2009}), \bibinfo{pages}{333--389}.
\newblock


\bibitem[Rose et~al\mbox{.}(2010)]%
        {rose2010automatic}
\bibfield{author}{\bibinfo{person}{Stuart Rose}, \bibinfo{person}{Dave Engel},
  \bibinfo{person}{Nick Cramer}, {and} \bibinfo{person}{Wendy Cowley}.}
  \bibinfo{year}{2010}\natexlab{}.
\newblock \showarticletitle{Automatic keyword extraction from individual
  documents}.
\newblock \bibinfo{journal}{\emph{Text mining: applications and theory}}
  (\bibinfo{year}{2010}), \bibinfo{pages}{1--20}.
\newblock


\bibitem[Santhanam et~al\mbox{.}(2022)]%
        {santhanam2022colbertv2}
\bibfield{author}{\bibinfo{person}{Keshav Santhanam}, \bibinfo{person}{Omar
  Khattab}, \bibinfo{person}{Jon Saad-Falcon}, \bibinfo{person}{Christopher
  Potts}, {and} \bibinfo{person}{Matei Zaharia}.}
  \bibinfo{year}{2022}\natexlab{}.
\newblock \showarticletitle{ColBERTv2: Effective and Efficient Retrieval via
  Lightweight Late Interaction}. In \bibinfo{booktitle}{\emph{Proceedings of
  the 2022 Conference of the North American Chapter of the Association for
  Computational Linguistics: Human Language Technologies}}.
  \bibinfo{pages}{3715--3734}.
\newblock


\bibitem[Shen et~al\mbox{.}(2014)]%
        {shen2014latent}
\bibfield{author}{\bibinfo{person}{Yelong Shen}, \bibinfo{person}{Xiaodong He},
  \bibinfo{person}{Jianfeng Gao}, \bibinfo{person}{Li Deng}, {and}
  \bibinfo{person}{Gr{\'e}goire Mesnil}.} \bibinfo{year}{2014}\natexlab{}.
\newblock \showarticletitle{A latent semantic model with convolutional-pooling
  structure for information retrieval}. In
  \bibinfo{booktitle}{\emph{Proceedings of the 23rd ACM international
  conference on conference on information and knowledge management}}.
  \bibinfo{pages}{101--110}.
\newblock


\bibitem[Tang et~al\mbox{.}(2021)]%
        {tang2021improving}
\bibfield{author}{\bibinfo{person}{Hongyin Tang}, \bibinfo{person}{Xingwu Sun},
  \bibinfo{person}{Beihong Jin}, \bibinfo{person}{Jingang Wang},
  \bibinfo{person}{Fuzheng Zhang}, {and} \bibinfo{person}{Wei Wu}.}
  \bibinfo{year}{2021}\natexlab{}.
\newblock \showarticletitle{Improving document representations by generating
  pseudo query embeddings for dense retrieval}.
\newblock \bibinfo{journal}{\emph{arXiv preprint arXiv:2105.03599}}
  (\bibinfo{year}{2021}).
\newblock


\bibitem[Vaswani et~al\mbox{.}(2017)]%
        {vaswani2017attention}
\bibfield{author}{\bibinfo{person}{Ashish Vaswani}, \bibinfo{person}{Noam
  Shazeer}, \bibinfo{person}{Niki Parmar}, \bibinfo{person}{Jakob Uszkoreit},
  \bibinfo{person}{Llion Jones}, \bibinfo{person}{Aidan~N Gomez},
  \bibinfo{person}{{\L}ukasz Kaiser}, {and} \bibinfo{person}{Illia
  Polosukhin}.} \bibinfo{year}{2017}\natexlab{}.
\newblock \showarticletitle{Attention is all you need}. In
  \bibinfo{booktitle}{\emph{Advances in neural information processing
  systems}}. \bibinfo{pages}{5998--6008}.
\newblock


\bibitem[Wan et~al\mbox{.}(2016)]%
        {wan2016deep}
\bibfield{author}{\bibinfo{person}{Shengxian Wan}, \bibinfo{person}{Yanyan
  Lan}, \bibinfo{person}{Jiafeng Guo}, \bibinfo{person}{Jun Xu},
  \bibinfo{person}{Liang Pang}, {and} \bibinfo{person}{Xueqi Cheng}.}
  \bibinfo{year}{2016}\natexlab{}.
\newblock \showarticletitle{A deep architecture for semantic matching with
  multiple positional sentence representations}. In
  \bibinfo{booktitle}{\emph{Proceedings of the AAAI Conference on Artificial
  Intelligence}}, Vol.~\bibinfo{volume}{30}.
\newblock


\bibitem[Wang et~al\mbox{.}(2021)]%
        {wang2021gpl}
\bibfield{author}{\bibinfo{person}{Kexin Wang}, \bibinfo{person}{Nandan
  Thakur}, \bibinfo{person}{Nils Reimers}, {and} \bibinfo{person}{Iryna
  Gurevych}.} \bibinfo{year}{2021}\natexlab{}.
\newblock \showarticletitle{Gpl: Generative pseudo labeling for unsupervised
  domain adaptation of dense retrieval}.
\newblock \bibinfo{journal}{\emph{arXiv preprint arXiv:2112.07577}}
  (\bibinfo{year}{2021}).
\newblock


\bibitem[Wang et~al\mbox{.}(2022)]%
        {wang2022simlm}
\bibfield{author}{\bibinfo{person}{Liang Wang}, \bibinfo{person}{Nan Yang},
  \bibinfo{person}{Xiaolong Huang}, \bibinfo{person}{Binxing Jiao},
  \bibinfo{person}{Linjun Yang}, \bibinfo{person}{Daxin Jiang},
  \bibinfo{person}{Rangan Majumder}, {and} \bibinfo{person}{Furu Wei}.}
  \bibinfo{year}{2022}\natexlab{}.
\newblock \showarticletitle{Simlm: Pre-training with representation bottleneck
  for dense passage retrieval}.
\newblock \bibinfo{journal}{\emph{arXiv preprint arXiv:2207.02578}}
  (\bibinfo{year}{2022}).
\newblock


\bibitem[Wu et~al\mbox{.}(2022a)]%
        {wu2022query}
\bibfield{author}{\bibinfo{person}{Xing Wu}, \bibinfo{person}{Guangyuan Ma},
  {and} \bibinfo{person}{Songlin Hu}.} \bibinfo{year}{2022}\natexlab{a}.
\newblock \showarticletitle{Query-as-context Pre-training for Dense Passage
  Retrieval}.
\newblock \bibinfo{journal}{\emph{arXiv preprint arXiv:2212.09598}}
  (\bibinfo{year}{2022}).
\newblock


\bibitem[Wu et~al\mbox{.}(2022b)]%
        {wu2022contextual}
\bibfield{author}{\bibinfo{person}{Xing Wu}, \bibinfo{person}{Guangyuan Ma},
  \bibinfo{person}{Meng Lin}, \bibinfo{person}{Zijia Lin},
  \bibinfo{person}{Zhongyuan Wang}, {and} \bibinfo{person}{Songlin Hu}.}
  \bibinfo{year}{2022}\natexlab{b}.
\newblock \showarticletitle{Contextual mask auto-encoder for dense passage
  retrieval}.
\newblock \bibinfo{journal}{\emph{arXiv preprint arXiv:2208.07670}}
  (\bibinfo{year}{2022}).
\newblock


\bibitem[Xiao et~al\mbox{.}(2023)]%
        {xiao2023social4rec}
\bibfield{author}{\bibinfo{person}{Xuanji Xiao}, \bibinfo{person}{Huaqiang
  Dai}, \bibinfo{person}{Qian Dong}, \bibinfo{person}{Shuzi Niu},
  \bibinfo{person}{Yuzhen Liu}, {and} \bibinfo{person}{Pei Liu}.}
  \bibinfo{year}{2023}\natexlab{}.
\newblock \showarticletitle{Social4Rec: Distilling User Preference from Social
  Graph for Video Recommendation in Tencent}.
\newblock \bibinfo{journal}{\emph{arXiv preprint arXiv:2302.09971}}
  (\bibinfo{year}{2023}).
\newblock


\bibitem[Xie et~al\mbox{.}(2023)]%
        {xie2023t2ranking}
\bibfield{author}{\bibinfo{person}{Xiaohui Xie}, \bibinfo{person}{Qian Dong},
  \bibinfo{person}{Bingning Wang}, \bibinfo{person}{Feiyang Lv},
  \bibinfo{person}{Ting Yao}, \bibinfo{person}{Weinan Gan},
  \bibinfo{person}{Zhijing Wu}, \bibinfo{person}{Xiangsheng Li},
  \bibinfo{person}{Haitao Li}, \bibinfo{person}{Yiqun Liu}, {et~al\mbox{.}}}
  \bibinfo{year}{2023}\natexlab{}.
\newblock \showarticletitle{T2Ranking: A large-scale Chinese Benchmark for
  Passage Ranking}.
\newblock \bibinfo{journal}{\emph{arXiv preprint arXiv:2304.03679}}
  (\bibinfo{year}{2023}).
\newblock


\bibitem[Xiong et~al\mbox{.}(2020)]%
        {xiong2020approximate}
\bibfield{author}{\bibinfo{person}{Lee Xiong}, \bibinfo{person}{Chenyan Xiong},
  \bibinfo{person}{Ye Li}, \bibinfo{person}{Kwok-Fung Tang},
  \bibinfo{person}{Jialin Liu}, \bibinfo{person}{Paul Bennett},
  \bibinfo{person}{Junaid Ahmed}, {and} \bibinfo{person}{Arnold Overwijk}.}
  \bibinfo{year}{2020}\natexlab{}.
\newblock \showarticletitle{Approximate nearest neighbor negative contrastive
  learning for dense text retrieval}.
\newblock \bibinfo{journal}{\emph{arXiv preprint arXiv:2007.00808}}
  (\bibinfo{year}{2020}).
\newblock


\bibitem[Yan et~al\mbox{.}(2021)]%
        {yan2021unified}
\bibfield{author}{\bibinfo{person}{Ming Yan}, \bibinfo{person}{Chenliang Li},
  \bibinfo{person}{Bin Bi}, \bibinfo{person}{Wei Wang}, {and}
  \bibinfo{person}{Songfang Huang}.} \bibinfo{year}{2021}\natexlab{}.
\newblock \showarticletitle{A Unified Pretraining Framework for Passage Ranking
  and Expansion}. In \bibinfo{booktitle}{\emph{Proceedings of the AAAI
  Conference on Artificial Intelligence}}, Vol.~\bibinfo{volume}{35}.
  \bibinfo{pages}{4555--4563}.
\newblock


\bibitem[Ye et~al\mbox{.}(2022)]%
        {ye2022fast}
\bibfield{author}{\bibinfo{person}{Wenwen Ye}, \bibinfo{person}{Yiding Liu},
  \bibinfo{person}{Lixin Zou}, \bibinfo{person}{Hengyi Cai},
  \bibinfo{person}{Suqi Cheng}, \bibinfo{person}{Shuaiqiang Wang}, {and}
  \bibinfo{person}{Dawei Yin}.} \bibinfo{year}{2022}\natexlab{}.
\newblock \showarticletitle{Fast Semantic Matching via Flexible Contextualized
  Interaction}. In \bibinfo{booktitle}{\emph{Proceedings of the Fifteenth ACM
  International Conference on Web Search and Data Mining}}.
  \bibinfo{pages}{1275--1283}.
\newblock


\bibitem[You et~al\mbox{.}(2019)]%
        {you2019large}
\bibfield{author}{\bibinfo{person}{Yang You}, \bibinfo{person}{Jing Li},
  \bibinfo{person}{Sashank Reddi}, \bibinfo{person}{Jonathan Hseu},
  \bibinfo{person}{Sanjiv Kumar}, \bibinfo{person}{Srinadh Bhojanapalli},
  \bibinfo{person}{Xiaodan Song}, \bibinfo{person}{James Demmel},
  \bibinfo{person}{Kurt Keutzer}, {and} \bibinfo{person}{Cho-Jui Hsieh}.}
  \bibinfo{year}{2019}\natexlab{}.
\newblock \showarticletitle{Large batch optimization for deep learning:
  Training bert in 76 minutes}.
\newblock \bibinfo{journal}{\emph{arXiv preprint arXiv:1904.00962}}
  (\bibinfo{year}{2019}).
\newblock


\bibitem[Zhou et~al\mbox{.}(2022)]%
        {zhou2022master}
\bibfield{author}{\bibinfo{person}{Kun Zhou}, \bibinfo{person}{Xiao Liu},
  \bibinfo{person}{Yeyun Gong}, \bibinfo{person}{Wayne~Xin Zhao},
  \bibinfo{person}{Daxin Jiang}, \bibinfo{person}{Nan Duan}, {and}
  \bibinfo{person}{Ji-Rong Wen}.} \bibinfo{year}{2022}\natexlab{}.
\newblock \showarticletitle{MASTER: Multi-task Pre-trained Bottlenecked Masked
  Autoencoders are Better Dense Retrievers}.
\newblock \bibinfo{journal}{\emph{arXiv preprint arXiv:2212.07841}}
  (\bibinfo{year}{2022}).
\newblock


\bibitem[Zou et~al\mbox{.}(2022)]%
        {zou2022pre}
\bibfield{author}{\bibinfo{person}{Lixin Zou}, \bibinfo{person}{Weixue Lu},
  \bibinfo{person}{Yiding Liu}, \bibinfo{person}{Hengyi Cai},
  \bibinfo{person}{Xiaokai Chu}, \bibinfo{person}{Dehong Ma},
  \bibinfo{person}{Daiting Shi}, \bibinfo{person}{Yu Sun},
  \bibinfo{person}{Zhicong Cheng}, \bibinfo{person}{Simiu Gu}, {et~al\mbox{.}}}
  \bibinfo{year}{2022}\natexlab{}.
\newblock \showarticletitle{Pre-trained language model-based retrieval and
  ranking for Web search}.
\newblock \bibinfo{journal}{\emph{ACM Transactions on the Web}}
  \bibinfo{volume}{17}, \bibinfo{number}{1} (\bibinfo{year}{2022}),
  \bibinfo{pages}{1--36}.
\newblock


\end{thebibliography}
\end{document}